\documentclass[aps,prb,preprint,groupedaddress]{revtex4-1}

\usepackage[pdftex]{graphicx}
\usepackage{multirow}

\begin{document}


\title{Raman spectrum and optical extinction of graphene buffer layers on the Si-face of 6H-SiC}


\author{A. Tiberj$^{1,2}$}
\author{J. R. Huntzinger$^{1,2}$}
\author{N. Camara$^{1,2,3,*}$}
\author{P. Godignon$^3$}
\author{J. Camassel$^{1,2}$}
\affiliation{$^1$Universit\'e Montpellier 2, Laboratoire Charles Coulomb UMR 5221, F-34095, Montpellier, France}
\affiliation{$^2$CNRS, Laboratoire Charles Coulomb UMR 5221, F-34095, Montpellier, France}
\affiliation{$^3$IMB-CNM-CSIC, Campus UAB Bellaterra, Barcelona 08193, Spain}
\altaffiliation{present address: University of Tours, CNRS, Laboratoire GREMAN UMR 7347, 16, rue Pierre et Marie Curie, 37071 Tours Cedex 2, France }

\author{}
\affiliation{}


\date{\today}

\begin{abstract}
The buffer layer has been analysed by combined micro-Raman and micro-transmission experiments. The epitaxial graphene growth on the (0001) Si face of 6H-SiC substrates was tuned to get a mixed surface at the early stage of graphitization with i) bare SiC, ii) buffer layer and iii) in some localized areas small monolayers flakes on top of the buffer layer. These unique samples enabled to measure the Raman spectrum of the buffer layer (close to the Raman spectrum of a carbon layer with a significant percentage of sp$^3$ bonds) and its corresponding relative extinction at 514.5~nm. The Raman spectrum of the buffer layer remains visible after the growth of one monolayer on top but, despite the relatively low absorption coefficient of graphene, the Raman intensity is strongly reduced (typically divided by 3). The buffer layer background will bias usual evaluations of the domain sizes based on the D/G integrated intensities ratio. Finally, several Raman maps show the excellent thickness uniformity and crystalline quality of the graphene monolayers and that they are subjected to a non uniform compressive strain with  values comprised between $-0.60\% < \varepsilon <-0.42\%$.
\end{abstract}


\maketitle

\section{INTRODUCTION}
A major concern for epitaxial graphene on SiC (EG/SiC)  is the interface structure between the first graphene layer and the underlying SiC substrate. For usual (0001) SiC wafer orientation, it depends strongly on the growth conditions and orientation (Si vs C face) on which graphene is grown. For a recent review, see Ref.~\onlinecite{0022-3727-45-15-154001}. On the Si face, large homogeneous graphene monolayers (MLs) and bilayers (BLs) can be obtained on top of a $6\sqrt3 \times 6\sqrt 3R30$ SiC surface reconstruction (noted hereafter 6R3)\cite{Berger:2004bh,Sutter:2009dq,PhysRevB.78.245403,Emtsev:2009nx,PhysRevLett.102.106104,PhysRevLett.100.176802,PhysRevLett.105.085502,PhysRevB.77.235412}. These graphene planes are Bernal (AB) stacked and the interface between the first graphene plane and the SiC is made of an intermediate C-rich layer  (called buffer layer) which has covalent bonds with Si atoms of the substrate \cite{PhysRevLett.100.176802,PhysRevLett.105.085502,PhysRevB.77.235412}. On the C face, the situation is completely different. There is no buffer layer anymore. The interaction between the first graphene layer and the C atoms of the SiC-C face is reduced. Instead of a single 6R3, two different pristine surface reconstructions may exist below the first graphene layer: $(2 \times 2)_c$ and $(3\times3)$. Moreover, the graphene layers may have several orientations on top of each surface reconstruction \cite{Hoster:1997tg,PhysRevB.77.155303,PhysRevB.78.153412,PhysRevB.80.235429}. Finally, the growth rate  on the C face is higher, which makes the growth much more difficult to control at the full wafer scale.\\ 

Coming back to the Si face, the main issue for electronic device applications has long been the disappointingly low mobility of carriers (usually few thousands cm$^2$.V$^{-1}$.s$^{-1}$) compared to exfoliated graphene or EG grown on the C face (between 10000 to 27000 cm$^2$.V$^{-1}$.s$^{-1}$) \cite{Berger:2006qf,wu:223108}. This was explained by the presence of the buffer layer acting as a primary source of carrier doping and scattering\cite{PhysRevLett.99.126805,0022-3727-43-37-374009}. The corresponding (residual) n-type doping is around $10^{13}~\textrm{cm}^{-2}$, pinning the Fermi level energy at about 420 meV above the Dirac point. As a consequence, to improve the transport properties, it is needed to avoid or remove the buffer layer. This has been done by passivating the Si dangling bonds either by post-growth hydrogen annealing\cite{Virojanadara2010L4,0022-3727-43-37-374010,PhysRevLett.103.246804,0022-3727-43-37-374009,PhysRevB.84.125449,0953-8984-21-13-134016} or by growing graphene with propane CVD with H$_2$ gas vector\cite{10.1063/1.3503972}. In both cases, controlling the degree of passivation by fast, non-destructive, optical techniques is important to improve the results.\\

Unfortunately, to date, the crystalline and electronic structure of the buffer layer has only been studied by XPS, LEED and ARPES measurements \cite{PhysRevB.77.155303,Virojanadara2010L4,0022-3727-43-37-374010,PhysRevLett.103.246804,0022-3727-43-37-374009,PhysRevB.84.125449,0953-8984-21-13-134016}. They confirmed the semiconducting character as theoretically predicted\cite{PhysRevLett.100.176802,PhysRevB.77.235412,PhysRevLett.105.085502}, but the optical response remains to be investigated. Up to now, despite the fact that optical experiments ``easier'' to perform, the Raman spectrum of this buffer layer has only been evidenced once by performing depolarized Raman spectroscopy experiments\cite{PhysRevLett.101.156801}. The reduction of the intensity of the 2nd-order SiC Raman spectrum enabled direct observation of the usual D and G Raman bands of graphitic materials. This technique was successful to evidence  the large structural compressive strain of EG monolayers grown on the Si face. However, half of the G band intensity is lost when using an analyser and the buffer layer spectra were still perturbed by the remaining part of the 2nd-order Raman spectrum of the SiC substrate.\\

In this work, we focus on the optical response of the buffer layer and, especially, on the comparison of its optical absorption and Raman response. We first describe how the graphene growth was tuned at the early stage of graphitization to get a mixed surface with the coexistence of bare SiC, buffer layer areas and small monolayers flakes on top of the buffer layer. Then, the Raman spectra and relative extinctions of the buffer layer and the monolayer on top of the buffer layer are detailed and compared. Finally, micro-Raman and micro-transmission maps of these samples are presented.

\section{EXPERIMENTS}
Graphene was grown on top of  $0.8\times0.8$ cm$^2$ semi-insulating, on-axis, 6H-SiC(0001) substrates using a commercial RF-induction furnace from Jipelec\cite{PhysRevB.80.125410}.  Before sublimation, the samples were cleaned using standard clean-room compatible RCA treatments. To focus on the early stage of growth, we performed graphitization under Ar close to atmospheric pressure. To achieve two different stages of graphitization, the samples were heated for 20 min at 1800 and 1850$^\circ$C. They were first characterized by conventional techniques, like optical microscopy, scanning electron microscopy (SEM) and atomic force microscopy (AFM). Then, combined micro-Raman spectroscopy and micro-transmission measurements were done, using a Jobin Yvon-Horiba T64000 spectrometer in the confocal mode fitted with a $\times100$ microscope objective. The 514.5~nm line of an Ar ion-laser was used for excitation. The spot diameter was 1 $\mu$m, with 1-mW incident power below the objective. A more detailed description of the experimental configuration can be found in Refs.~\onlinecite{PhysRevB.80.125410,Camara_2010}\\

\section{RESULTS AND DISCUSSION}
\subsection{Buffer layer}
\begin{figure}[t]
\begin{minipage}[l]{0.45\columnwidth}
\includegraphics[height=6cm]{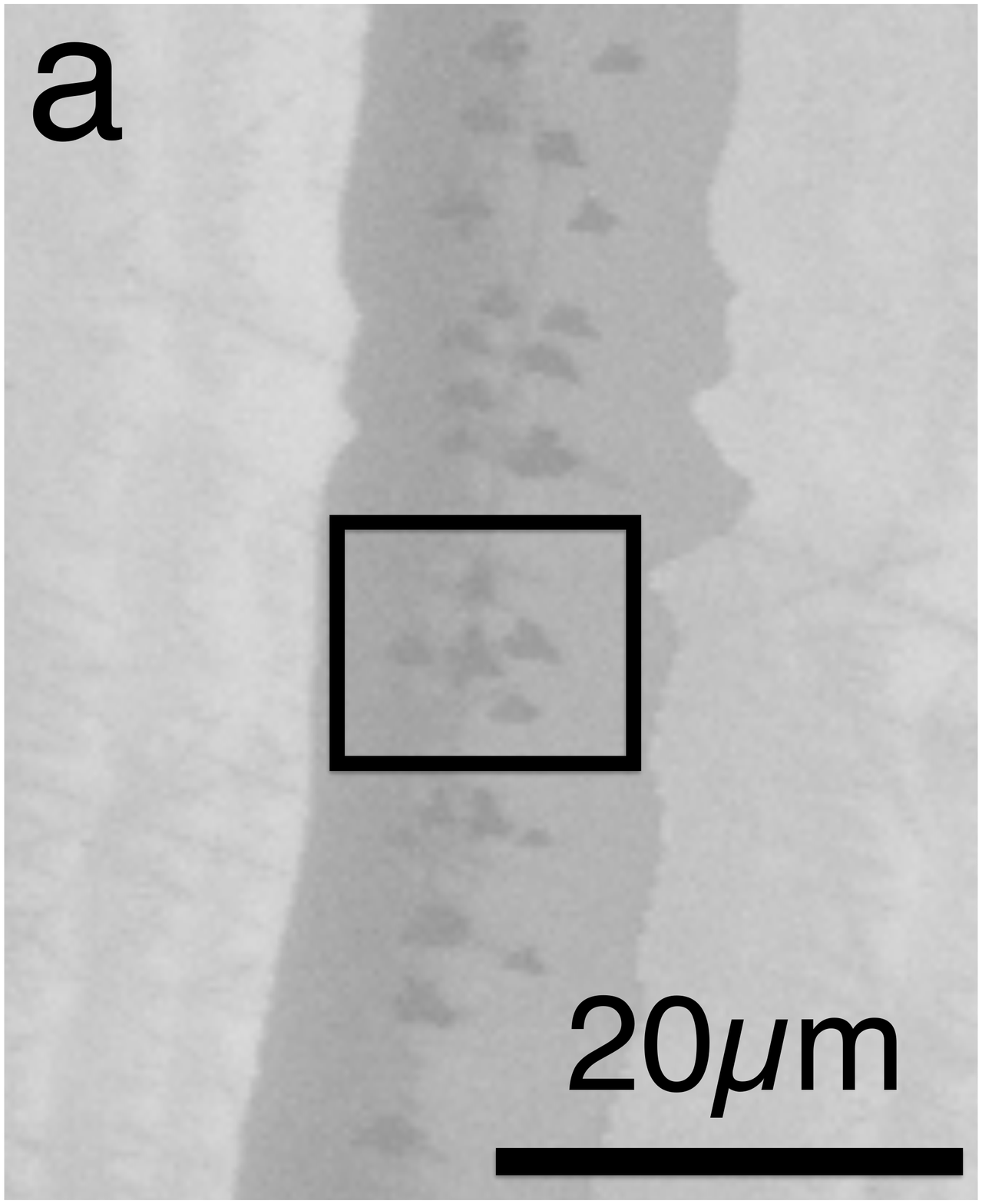}
\end{minipage}
\hspace{0cm}
\begin{minipage}[l]{0.45\columnwidth}
\includegraphics[height=6cm]{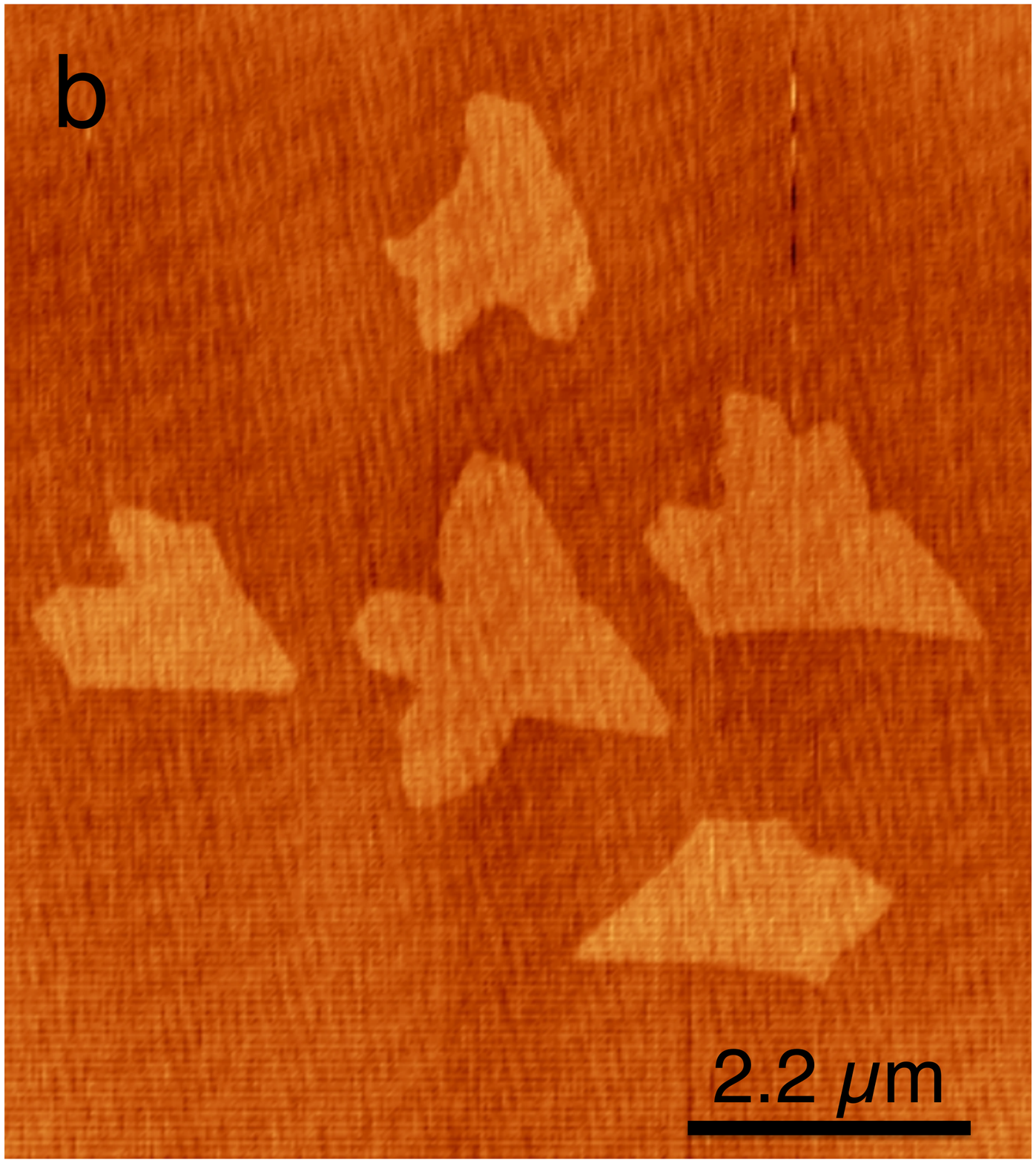}
\end{minipage}
\caption{\label{SEM_AFM}a) SEM pictures of a partly graphitized surface. Three areas can be distinguished. The light grey in the left and right hand-side of the image corresponds to bare SiC, the mid grey area corresponds to a large SiC surface fully covered by the buffer layer ($\sim10 \mu m$ large step). On top of this buffer layer, darker flakes appear and correspond to graphene monolayers on top of the buffer layer. A typical example, corresponding to the square area is enlighted in Fig.~\ref{SEM_AFM}b. b) AFM zoom of five monolayers flakes selected in the square area of Fig.~\ref{SEM_AFM}a. The step height between the buffer and the flakes ranges from 3 to 4 \AA, which strongly suggests that the flakes are graphene monolayers.
 }
\end{figure}

On the first sample, grown at 1800$^\circ$C for 20 min, the graphitization was not at all homogenous. This is shown in the SEM and AFM pictures of Fig.~\ref{SEM_AFM}. Consider, first, Fig.~\ref{SEM_AFM}a. On the right and left parts, unreconstructed terraces are still made of bare SiC. In between a large step, fully reconstructed, is covered by the buffer layer. On top of this buffer layer, small graphene monolayer flakes start developing. In the SEM picture, the bare SiC areas appear as light grey, while the SiC part covered by the buffer layer are shown as mid grey. On top of the buffer layer, the small (darker) objects are monolayer flakes. For convenience, five of them have been enlighted in the square. In Fig.~\ref{SEM_AFM}b, they have been probed by AFM, which reveals a topological height difference between 3 and 4 \AA, without any trace of wrinkles. This suggests that these flakes correspond to graphene MLs on top of the buffer layer.\\

\begin{figure}[t]
\includegraphics[width=0.95\columnwidth]{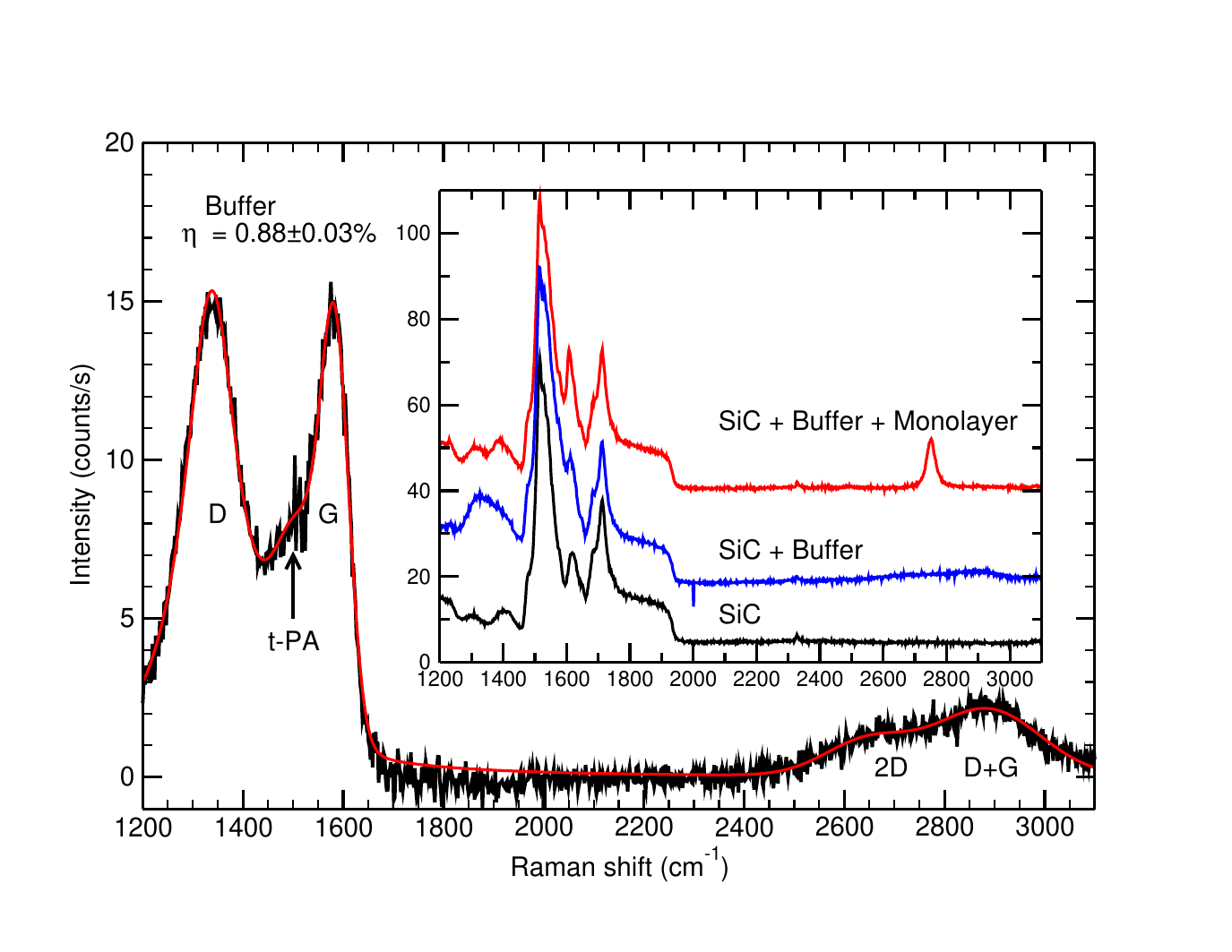}
\caption{\label{Raman_buffer}Raman spectrum of the buffer layer obtained after SiC spectrum substraction. Notice the large and broad D and G bands. Notice also the weak trans-polyacethylene contribution noted t-PA. This spectrum does not exhibit any sharp 2D-band, but only broad 2D and D+G overtones extending from 2400 to 3000 cm$^{-1}$. The relative extinction value $\eta=0.88\%$ is extracted from micro-transmission measurements. The red line is a fit calculated with parameter values listed in Table~\ref{table_buffer}. Insert: raw Raman spectra of bare SiC, the buffer layer and a monolayer graphene flake on top of the buffer layer.}
\end{figure}

To confirm these results, Raman spectra were collected on these three different areas. Raw spectra are shown in the insert of Fig.~\ref{Raman_buffer}. They confirm that the light grey areas seen in Fig.~\ref{SEM_AFM}a correspond to bare SiC, with only the typical 2nd-order Raman spectrum of 6H-SiC\cite{PhysRevB.59.7282}. The buffer layer spectrum, collected on the large reconstructed terrace (mid-grey area in Fig.~\ref{SEM_AFM}a) is displayed in blue. At first glance, it seems very close to the SiC one with, however, some differences. The main (visible) differences are: i) a broad band around 1300 cm$^{-1}$ ii) a high energy broadening of the  strong 2nd-order SiC Raman feature at 1550 cm$^{-1}$ and iii) the apparition of two broad and weak bands between 2800 and 3000 cm$^{-1}$. Finally, in red, the small flakes exhibit the typical fingerprint of graphene with a narrow 2D mode and G mode overimposed on the 2nd-order SiC overtones. From these spectra, after proper substraction of the SiC Raman fingerprint, the Raman spectra of the buffer and the monolayer on top of the buffer can be obtained. They are shown in Fig.~\ref{Raman_buffer} (bottom spectrum) and Fig.~\ref{Raman_mono}, respectively.\\
Consider first the Raman spectrum of the buffer layer. It has all the hallmarks highly functionalized graphene or graphene oxide spectra\cite{Englert:2011fk}. It displays mainly two large bands, with a broad D-like feature at 1337 cm$^{-1}$ (FWHM = 135 cm$^{-1}$) and a G-like band at 1584 cm$^{-1}$ (FWHM = 65 cm$^{-1}$). In between is a small hump at 1515 cm$^{-1}$. This additional feature has already been observed and ascribed to trans-polyacetylene (t-PA) chains formed at the grain boundaries of nano-crystalline diamonds\cite{Englert:2011fk,PhysRevB.63.121405}. The spectrum does not exhibit any sharp 2D-band, but broad 2D and D+G overtones extending from 2400 to 3000 cm$^{-1}$. The large D band and D/G ratio indicate a large number of crystalline defects with domain size L$_D$ of the order of 1~nm\cite{PhysRevB.82.125429}. The Raman spectrum of the buffer layer is indeed very close to the one observed in Ref.~\onlinecite{PhysRevB.82.125429} for graphene after ion implantation with 10$^{14}$ Ar$^+$-ion cm$^{-2}$. It confirms the usual picture of an initial coverage by a C-rich buffer layer made of graphitic-like clusters bonded to the SiC surface by a significant percentage of sp$^3$ C atoms\cite{0022-3727-43-37-374009}. This spectrum has been modeled by using a 5-oscillators fit, with parameter values listed in Table~\ref{table_buffer}. The result is shown as full line in Fig.~\ref{Raman_buffer}. In order to check the reproducibility, 4 spectra were collected at different points of this sample; 4 other spectra were collected on the sample grown at 1850$^\circ$C using the same acquisition conditions (3x150s). All the parameter values used to fit these 8 different spectra are compared in Table~\ref{table_buffer} and show that the 50$^\circ$C increase do not change significantly the Raman fingerprint of the buffer layer. Finally these values are also compared to the ones extracted from the Raman mapping of Fig.~\ref{Raman_maps}. In this case, the standard deviations are higher because the shorter acquisition time of $2\times 10$~s decreases the signal to noise ratio.

\begin{turnpage}
\begin{table}
 \caption{\label{table_buffer} Parameters used to fit the Raman spectrum of the buffer layer shown in Fig. \ref{Raman_buffer}. All the bands were computed with a Gaussian line shape except for the D band which was calculated with a Lorentzian line shape. $A$ (counts.s$^{-1}$.cm$^{-1}$),  $\omega$ (cm$^{-1}$) and $\Gamma$ (cm$^{-1}$) are respectively the integrated intensity, the position and the FWHM of the band. These parameters are then compared to the average and the standard deviations obtained from 8 Raman spectra of the buffer layer on the two samples studied in this article, and to the parameters used to fit the buffer spectra measured during the Raman mapping shown in Fig.~\ref{Raman_maps}. The acquisition time was then of 2x10s (instead of 3x150s for the 8 former spectra) with a poorer signal to noise ratio.}
 \begin{ruledtabular}
 \begin{tabular}{c|c c c | c c c | c c c | c c c | c c c}
\multirow{2}{*}{Spectra} &\multicolumn{3}{c|}{D} & \multicolumn{3}{c|}{t-PA} & \multicolumn{3}{c|}{G} & \multicolumn{3}{c|}{2D} & \multicolumn{3}{c}{D+G} \\

&$A_D$ & $\omega_D$ & $\Gamma_D$ & $A_{tpa}$ & $\omega_{tpa}$ & $\Gamma_{tpa}$ & $A_G$ & $\omega_G$ & $\Gamma_G$ & $A_{2D}$ & $\omega_{2D}$ & $\Gamma_{2D}$ & $A_{D+G}$ & $\omega_{D+G}$ & $\Gamma_{D+G}$ \\

\hline
\multicolumn{16}{c}{\textbf{Buffer layer grown at 1800$^\circ$C }} \\
\hline
Fig.~\ref{Raman_buffer} fit & 3260 & 1337 & 135 & 797 & 1515 & 123 & 771 & 1584 & 65 & 237 & 2647 & 203 & 564 & 2884 & 251\\
\hline

average of 4 spectra & 3187 & 1337 & 129 & 661 & 1506 & 130 & 929 & 1582 & 69 & 206 & 2653 & 181 & 505 & 2882 & 237\\
standard deviations & 53 & 1 & 7 & 99 & 7 & 6 & 119 & 2 & 3 & 40 & 15 & 22 & 91 & 4 & 34\\

\hline
Mean values of Fig.~\ref{Raman_maps}&3278 & 1335 & 142 & 353 & 1483 & 84 & 995 & 1578 & 70 & 836 & 2646 & 300 & 840 & 2951 & 296 \\
standard deviations& 204 & 2 & 7 & 135 & 14 & 26 & 114 & 3 & 5 & 133 & 56 & 5 & 182 & 27 & 23 \\
\hline
\multicolumn{16}{c}{\textbf{Buffer layer grown at 1850$^\circ$C }} \\
\hline
average of 4 spectra & 3430 & 1335.7 & 124 & 796 & 1510 & 131 & 1081 & 1580 & 68 & 223 & 2661 & 166 & 583 & 2881 & 227\\
standard deviations & 67 & 0.4 & 2 & 67 & 4 & 3 & 45 & 1 & 1 & 27 & 2 & 9 & 26 & 4 & 10\\

 \end{tabular}
 \end{ruledtabular}
 \end{table}
 \end{turnpage}

Before closing this section, one should notice that the Raman integrated intensity of the buffer layer is extremely strong compared to the one collected from graphene monolayer. It should also be compared to the relative extinction induced at 2.41 eV (5145 \AA) by the presence of this buffer layer. Averaging over a large number of micro-transmission data points (from point by point and mapping measurements), we find a typical value $\eta=0.88\pm 0.03\%$ that corresponds approximately to $\frac{2}{3}$ of the relative extinction value for a monolayer of graphene on SiC (1.23\%)\cite{PhysRevB.80.125410}. From a qualitative point of view (and considering that only the sp$^2$ bonds contribute to the optical conductivity in this energy range) this suggests that the buffer layer is made of $\frac{2}{3}$ of sp$^2$ bonded C atoms. The remaining $\frac{1}{3}$ of the C atoms should then form sp$^3$ bonds with the Si atoms of the SiC substrate, which would not contribute (or only weakly) to the optical extinction. This ratio between sp$^2$ and sp$^3$ bonds is in good agreement with previous structural studies\cite{0022-3727-43-37-374009}.
 
\subsection{Monolayer graphene on top of the buffer layer}

\begin{figure}
\includegraphics[width=0.95\columnwidth]{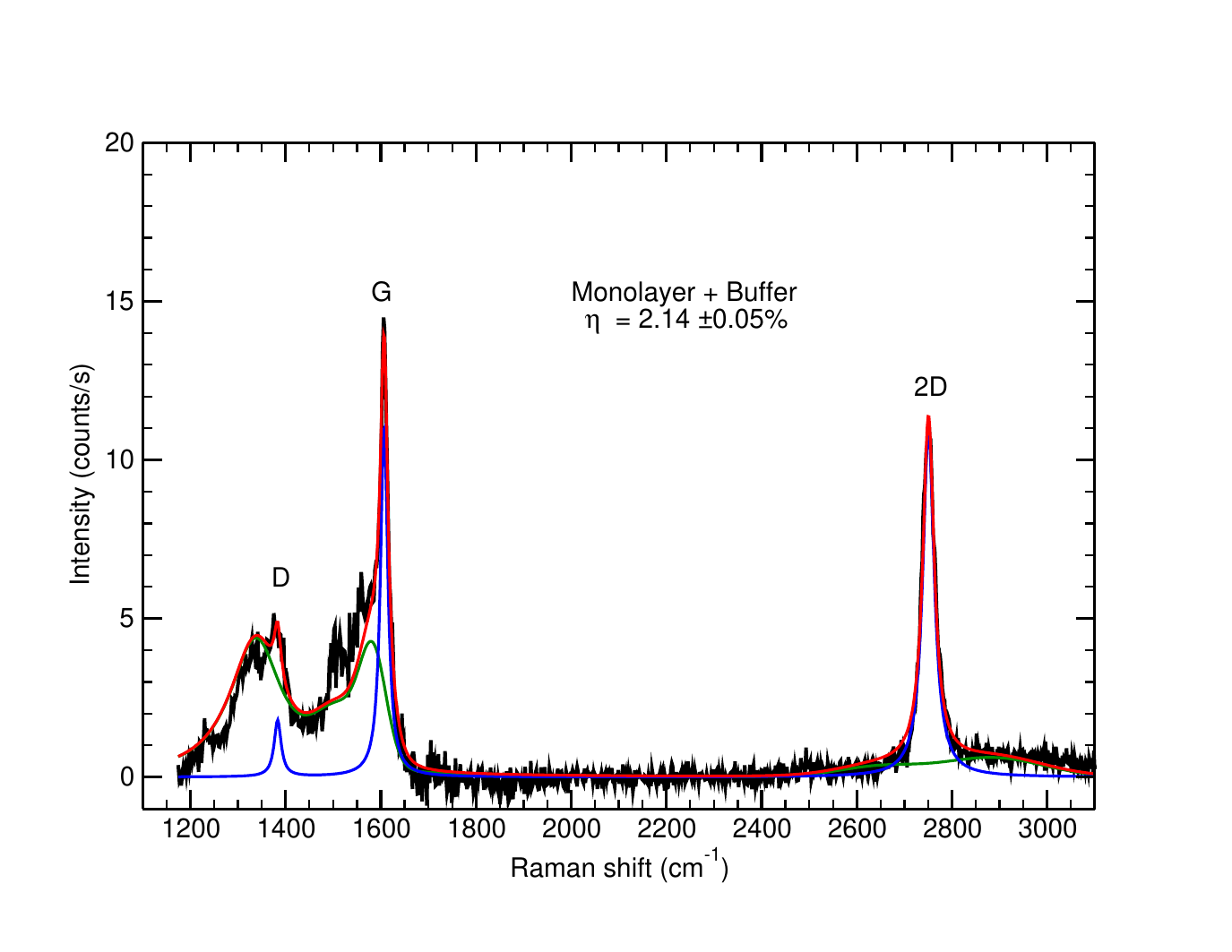}
\caption{\label{Raman_mono}Raman spectrum collected for a graphene monolayer grown on top of the buffer layer. The corresponding relative extinction is $\eta = 2.14 \pm 0.05 \%$. This spectrum was fitted (red line) by adding two contributions. The buffer layer contribution (green line) is calculated by multiplying by 0.29 the Raman fingerprint of the buffer layer shown in Fig.~\ref{Raman_buffer}. The monolayer contribution (blue line) is calculated with three lorenzian functions to fit the sharp D, G and 2D bands. We ascribed the additional D band arbitrarily to the graphene monolayer. It is indeed impossible to discriminate if it comes really from defects in the monolayer or from crystalline modification in the buffer layer.}
\end{figure}

Let us now consider the small flakes shown in Fig.~\ref{SEM_AFM}b on top of the buffer layer. As already said, the striking difference with respect to the buffer layer alone is that the Raman spectra shown in Fig.~\ref{Raman_buffer} (insert) and Fig.~\ref{Raman_mono} reveal now the sharp 2D-band of graphene at 2751 cm$^{-1}$ with FWHM = 28.9 cm$^{-1}$. A narrow G-band appears also at 1607cm$^{-1}$ with FWHM = 18.6 cm$^{-1}$. Below these true graphene features, remain the broad D and G bands previously found on the buffer layer (simply attenuated). 
This is best seen in Fig.~\ref{Raman_mono} in which two contributions have been used to fit the experimental results. One comes from the graphene monolayer. The second one comes from the buffer layer underneath. For the graphene monolayer, we use three lorenzian functions for the D, G and 2D bands, with parameter values listed in Table~\ref{table_monolayer}. For the buffer layer, we take the Raman fingerprint of Fig.~\ref{Raman_buffer} multiplied by a proportionality factor of 0.29. The result of the fit is displayed as red line in Fig.~\ref{Raman_mono}. We also plotted as blue and green lines, the separate contributions coming from the graphene flake and the buffer layer underneath, respectively. Notice that the small width of the lorentzian 2D-band is not new. It is typical of graphene MLs or twisted multilayers on top of SiC or SiO$_2$/Si\cite{PhysRevB.80.125410,PhysRevB.79.195417,PhysRevB.77.235403}. The normalized integrated intensity of the G band that corresponds to the monolayer contribution (fitted in blue) is also not new. We recover a value close to 0.03 which corresponds well to MLs \cite{PhysRevB.80.125410,21801347}. Finally, we ascribe the sharp (additional) D band arbitrarily to ML graphene since it is impossible to discriminate if it comes really from defects in the monolayer or from crystalline modifications in the buffer layer. 

\begin{table}
 \caption{\label{table_monolayer} Parameters used to fit the Raman spectrum of the graphene monolayer on top of the buffer layer shown in Fig. \ref{Raman_mono}. The buffer ratio corresponds to the proportionality factor by which the buffer fit function is multiplied to reproduce the attenuated buffer Raman spectrum that is still visible. The parameters used to fit the HOPG spectrum is also displayed since it is used as frequency and intensity reference. The G band integrated intensity of the monolayer normalized by the HOPG one gives a value of 0.038 confirming that these flakes are monolayers. Finally we also show the average and standard deviations of the parameters used to fit the five monolayers flakes in the Raman mapping shown in Fig.~\ref{Raman_maps}}
 \begin{ruledtabular}
 \begin{tabular}{c c c c c c c c c c c }
\multirow{2}{*}{Spectra} &Buffer ratio & \multicolumn{3}{c}{D} & \multicolumn{3}{c}{G} & \multicolumn{3}{c}{2D} \\

& &$A_D$ & $\omega_D$ & $\Gamma_D$ & $A_G$ & $\omega_G$ & $\Gamma_G$ & $A_{2D}$ & $\omega_{2D}$ & $\Gamma_{2D}$  \\
\hline
Fig.~\ref{Raman_mono} fit& 0.29 & 49.4 & 1383 & 17.6 & 324 & 1607 & 18.6 & 498 & 2751 & 28.9 \\
\hline
Mean values of Fig.~\ref{Raman_maps}& 0.37 & 48.8 & 1385 & 15.3 & 263 & 1608 & 15.9 & 477 & 2752 & 37.5\\
standard deviations& 0.12 & 28.3 & 8.3 & 9.5 & 66.7 &  1.5 & 3.2 & 64.0 & 3.2 & 11.8\\
\hline
\multirow{2}{*}{HOPG} &  &  & & & \multirow{2}{*}{8598} & \multirow{2}{*}{1581.5} & \multirow{2}{*}{15.6} & 2724 & 2684 & 39.5\\ 
 &  &  & & &  &  &  & 7065 & 2725 & 32.6 \\ 
 \end{tabular}
 \end{ruledtabular}
 \end{table}

To complement these results, we measured  the relative extinction of the whole stack: buffer layer plus ML graphene flake. We found $\eta = 2.14 \pm 0.05 \%$ compared to bare SiC and $\eta=1.27 \pm 0.03\%$ compared to the buffer layer. This is in excellent agreement with previous results collected for ML graphene on the C-face\cite{PhysRevB.80.125410,Camara_2010}. Simply, with respect to the C-face, an offset value of $0.88\pm 0.03\%$ has to be taken into account to consider the absorption due to the buffer layer. \\

Finally, it is interesting to consider the two different behaviors of the graphene buffer layer noticed in transmission and Raman scattering. Indeed, while the absorbance are simply additive, the buffer layer Raman intensity is divided by 3 when covered by a graphene ML. This fact cannot be explained by the ML graphene attenuation of the Raman signal since ML graphene flake absorbs only 1.2 to 1.4\% of the light on a SiC substrate. 
The strong integrated intensity of the buffer signal can be related to bond 
polarizabilities disorder \cite{benassi95}. This kind of disorder can be 
efficient even in the perfectly periodic 6R3 structure, since the 
corresponding supercell is large.
A strong coupling between graphene and buffer has already been evidenced by
LEED or STM \cite{0022-3727-43-37-374009}, by the unusual band structure observed using
ARPES\cite{PhysRevLett.100.176802,PhysRevLett.105.085502} and by the low carrier mobility observed from magnetotransport
experiments\cite{Berger:2006qf,wu:223108,PhysRevLett.99.126805,0022-3727-43-37-374009} . This coupling between buffer and the well ordered graphene
could decrease the polarizabilities fluctuations, thereby reducing the buffer
signal. This phenomenon deserves more thorough theoretical investigations.

\subsection{Micro-Raman and micro-transmission mapping}
\begin{figure}
\includegraphics[width=0.5\columnwidth]{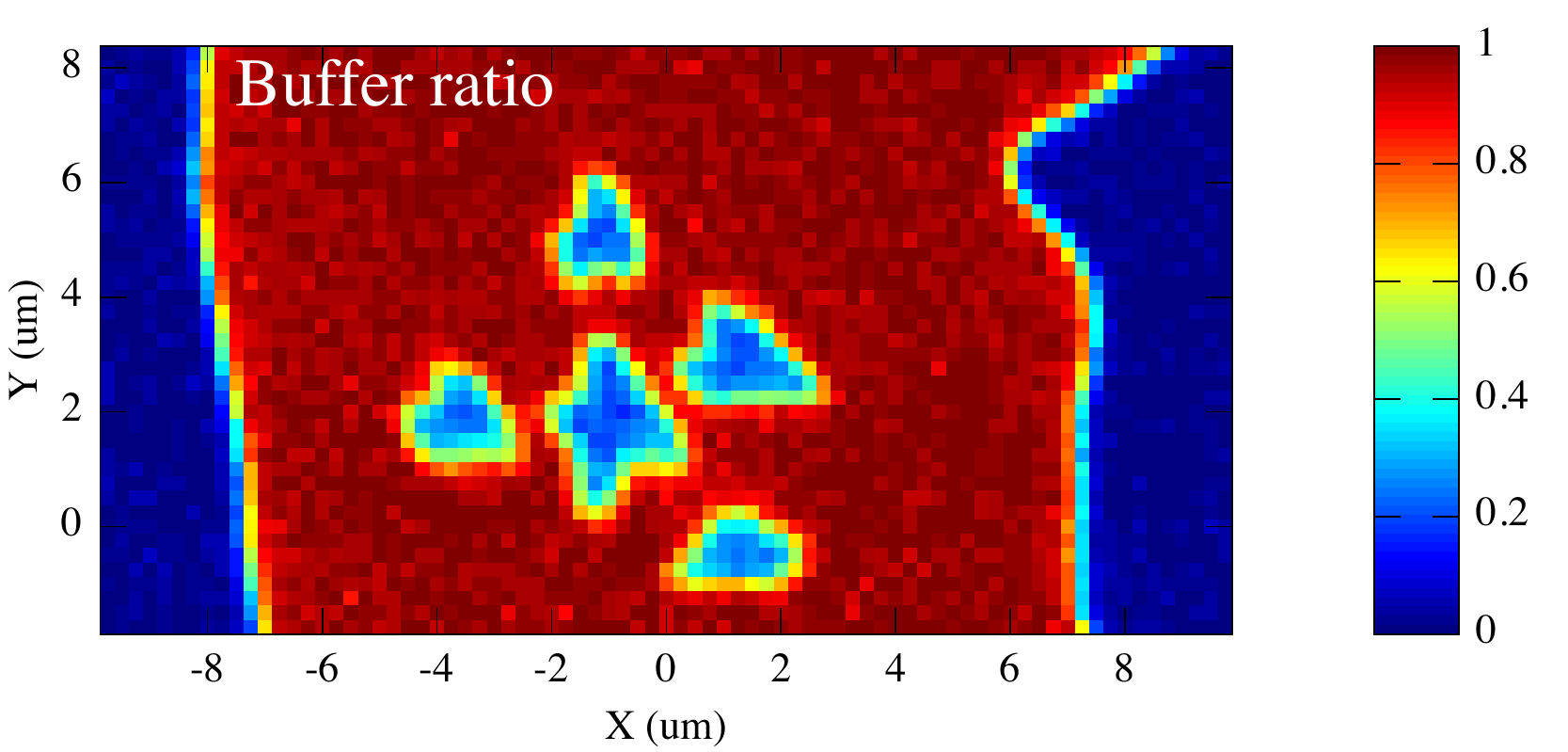}
\includegraphics[width=0.5\columnwidth]{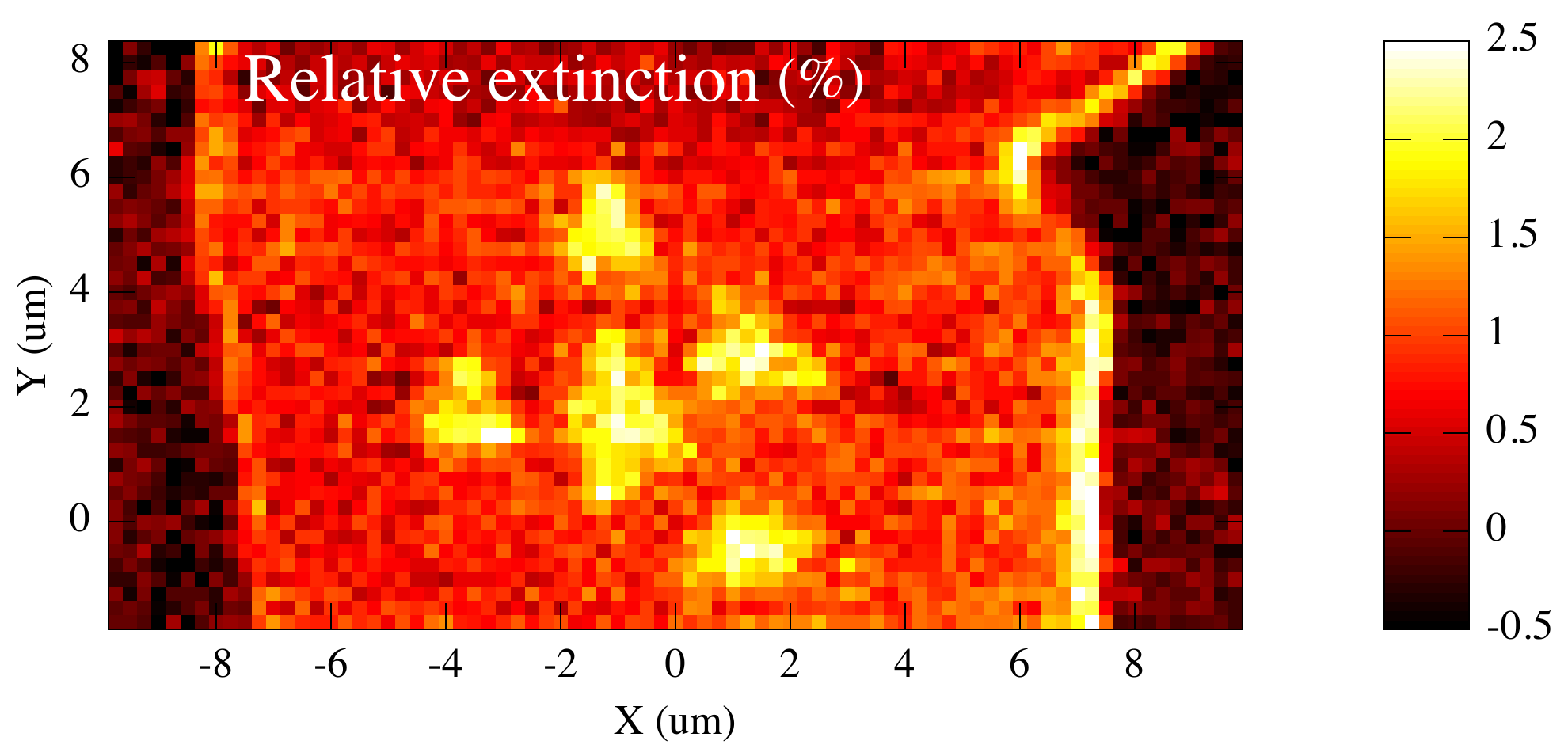}
\includegraphics[width=0.5\columnwidth]{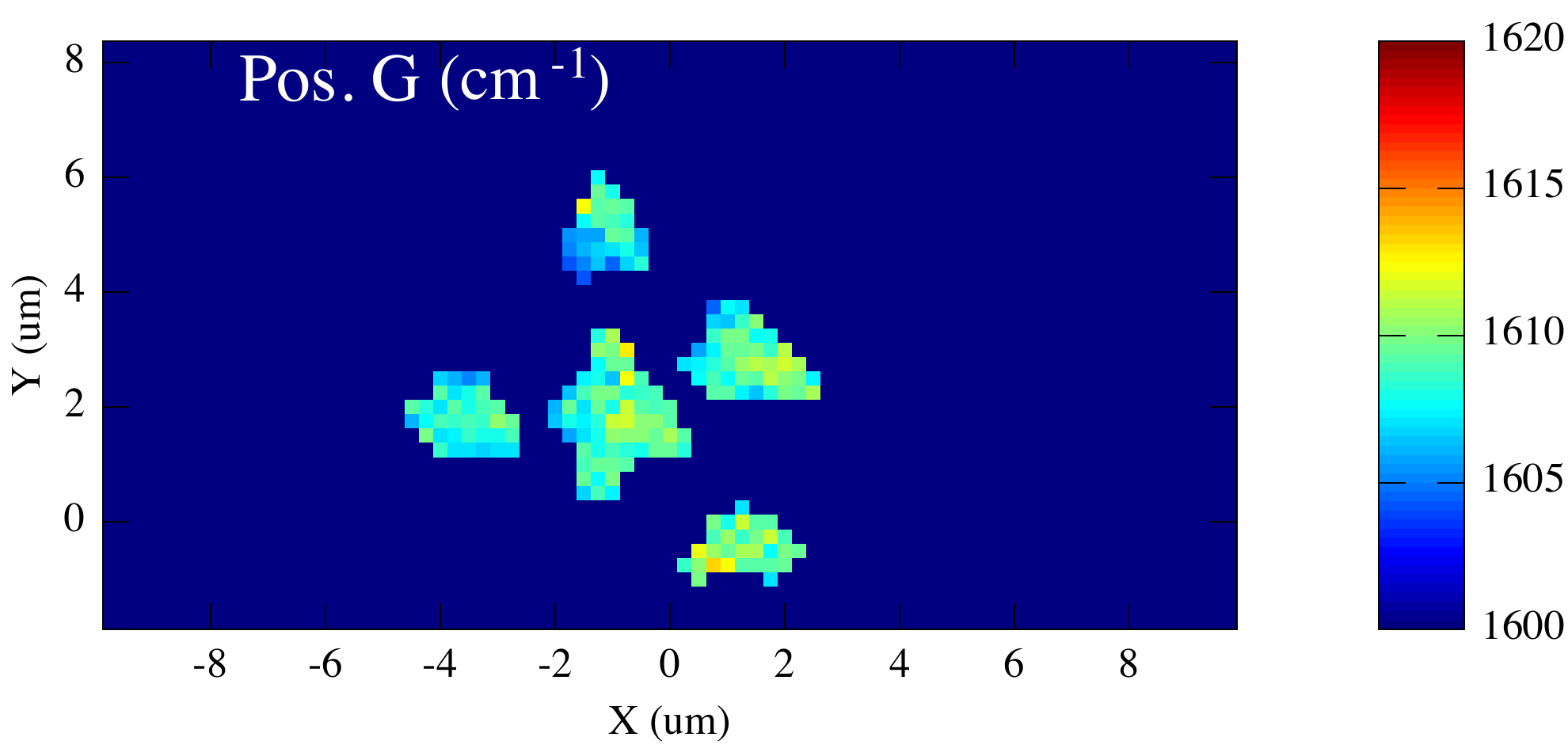}
\includegraphics[width=0.5\columnwidth]{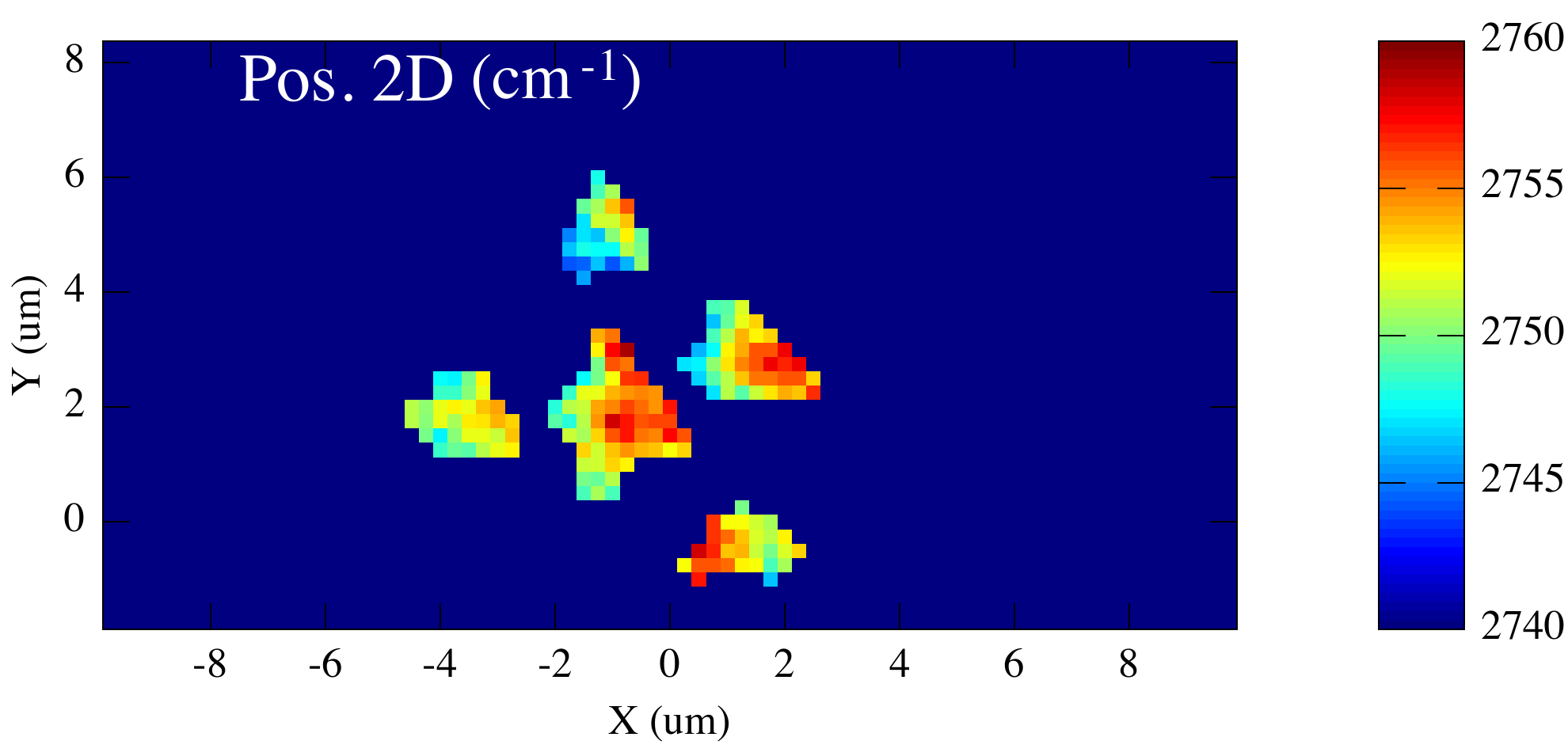}
\caption{\label{Raman_maps}Combined micro-Raman spectroscopy and micro-transmission mapping of the area shown in Fig.~\ref{SEM_AFM}. The buffer ratio, the relative extinction and the position of the G and 2D bands of the graphene are displayed. On the buffer map, we can clearly distinguished three areas: bare SiC substrate in blue with no buffer signal, the reconstructed SiC surface step covered by the buffer layer in red and finally the graphene monolayers flakes in blue with a Raman buffer signal divided by 3. The relative extinction shows also the values already reported around 0.8\% for the buffer and 2\% for the graphene on top of the buffer layer. Finally, the G and 2D bands of the graphene are strongly blueshifted. It indicates that these flakes are highly compressively stressed by the SiC substrate.
 }
\end{figure} 
To complement these results, the AFM area of Fig.~\ref{SEM_AFM}b has been imaged combining micro-Raman and micro-transmission measurements. $20\times10$~$\mu$m$^2$ maps with 0.25~$\mu$m steps were collected with an acquisition time of $2\times10$~s. The 3239 spectra were independently fitted. The extracted results are shown in Fig.~\ref{Raman_maps}. The first one shows the buffer relative intensity ratio in the Raman spectra on which three different areas can be clearly distinguished. The bare SiC surface with no buffer signal appears in blue with a buffer ratio of 0 on the left and right parts of the map. Second, the reconstructed SiC surface step covered by the buffer layer appears in red with a buffer ratio equal to 1. Finally, the graphene monolayer flakes on top of the buffer layer appears also in blue with a buffer signal divided by 3. The second map displays the corresponding relative extinctions from 0.8 to 0.9\% for the buffer layer and between 2.1 and 2.2\% for the monolayer on top of the buffer layer. The relative extinction values measured on the five different flakes, and the normalised integrated intensity of the G band (not shown), confirm the excellent thickness uniformity of the flakes. Notice that the relative extinction is higher on the edge of the reconstructed step, especially on the right part of the step. However, this higher relative extinction (close to 2\%) is not due to the presence of graphene monolayers on the step edge but comes from light scattering induced by the step that attenuates the transmitted intensity.\\

Finally, let us come to the evaluation of the residual (built-in) strain. The two last maps show that both the G and 2D bands are strongly blue shifted (by more than 70 cm$^{-1}$ for the 2D band) compared to epitaxial graphene grown on the C face of a SiC substrate\cite{PhysRevB.80.125410}. This is typical of the compressive strain usually observed on the Si face of SiC and comes from the difference between the thermal expansion coefficients of graphene and SiC\cite{PhysRevLett.101.156801}. If we assume that most of the shifts is due to this compressive strain, we can evaluate the strain using the relaxed position of graphene\cite{doi:10.1021/nl8031444} (1579.5 cm-1 for the G band and 2674 cm-1 for the 2D band using a 514.5~nm excitation wavelength) and the Gr\"uneisen parameters\cite{doi:10.1021/nl203359n} (1.8 for the G band and 2.6 for the 2D band). This gives an average strain $\varepsilon= -0.51\%$ (with 0.03\% standard deviation) from the G band position and $\varepsilon= -0.56\%$ (with 0.03\% standard deviation) from the 2D band. The good agreement between these two values confirm that most of the blueshift is due to the strong built-in (compressive) strain. Of course, there is an additional (minor) part due to the n-type doping induced by the buffer layer, but from these experiments it could not be evaluated. \\
 This large strain amplitude is confirmed by the absence of any wrinkles on the graphene flakes, as shown by AFM in Fig.~\ref{SEM_AFM}b. The formation of wrinkles is one of the strain relaxation mechanism commonly observed on the C-face of epitaxial graphene. Finally, we can also remark on the two maps of the G and 2D band position, that, despite the relatively small size of these flakes, this strain is not homogeneous: about 20 cm$^{-1}$ variations are observed on the 2D band position. The strain value are comprised between  $-0.60\% < \varepsilon <-0.42\%$. 

\subsection{Graphene grown at 1850$^\circ$C}
\begin{figure}
\includegraphics[width=0.6\columnwidth]{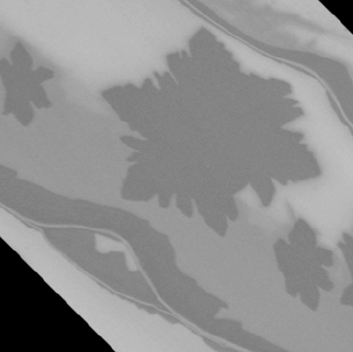}
\includegraphics[width=0.9\columnwidth]{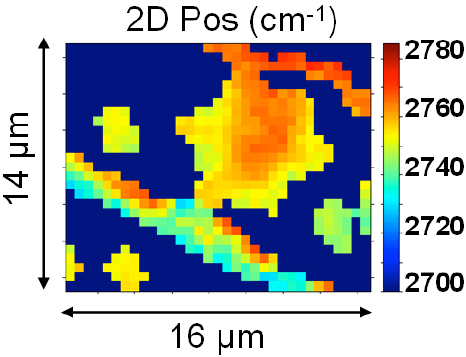}
\caption{\label{graphene_1850}SEM image of an advanced stage growth of monolayer of graphene on the Si-face at 1850$^\circ$C. Raman map of the 2D band position of the same area.
}
\end{figure}
On the second sample, grown at 1850$^\circ$C, a slightly more advanced state of graphitization was achieved. This resulted in larger homogeneous flakes, typically in the range of few microns. The SEM image shown in Fig.~\ref{graphene_1850} reveals that these graphene monolayer flakes grow anisotropically, with a flower-like shape. This is similar to the case of CVD graphene grown on metal\cite{Li:2009ly}. We also remark that a second growth mechanism starts at the edges of the reconstructed steps. Unfortunately, on this sample Raman spectroscopy did not reveal any bare SiC terrace. The whole sample surface was indeed covered by the buffer layer with exactly the same spectrum as the one shown previously in Fig.~\ref{Raman_buffer}. Micro-Raman and micro-transmission mapping (also shown in Fig.~\ref{graphene_1850}) have been performed on this area with 0.5$\mu$m steps and reveal thanks to the relative extinction and the normalised integrated intensity of the G band that the flower-like flakes are still homogenous and uniform monolayers of graphene. This is not true at the edges of terraces, where starts to grow a non controlled mix of mono, bi and trilayer. From the 2D-band position mapping, we can observe again that even in the larger flakes, the strain in the graphene is absolutely not homogeneously spread, higher in the center of the flake. This inhomogeneity induces a broadening of both G and 2D bands. It should also be taken into account for future device fabrication since it can alter all the processing steps and may affect the transport properties for further applications.\\
  
\section{Conclusion}
The Raman spectrum and relative extinction of the buffer layer have been measured. Its Raman spectrum corresponds to those observed for graphene layers with a significant percentage of sp$^3$ bonds. Its relative extinction $\eta=0.88\pm 0.03\%$ corresponds, approximately, to $\frac{2}{3}$ of the graphene one. The Raman spectrum and the relative extinction value are consistent with the known ratio of sp$^2$/sp$^3$ bonds of the buffer layer. We have also shown that the buffer layer Raman spectrum is still visible when graphene monolayer has been grown on top of it. The buffer layer background will bias usual evaluations of the domain sizes based on the D/G integrated intensities ratio. Indeed, it is impossible to discriminate wether the observed D band comes from the buffer layer or from the graphene monolayer. These experiments yield a fast and easy way of understanding and detecting the presence of the buffer layer and its crystalline reconstruction upon different annealing conditions (hydrogenation, oxidation for instance).

The strong coupling existing between the graphene monolayer and the buffer layer (previously evidenced by STM, LEED, ARPES and theoretical studies) is confirmed by our optical results. The first clue (which was already evidenced\cite{PhysRevLett.101.156801}) is the high strain of the monolayers grown on top of the buffer layer $-0.60\% < \varepsilon <-0.42\%$. The graphene layers are indeed pinned by the buffer layer which prevents wrinkles formation and therefore the strain relaxation. The second clue, our most striking result, is the strong Raman intensity reduction of the buffer layer due to the presence of the graphene overlayer.  The intensity is divided by 3. A plausible explanation is related to the strong coupling existing between graphene and the buffer layer. This coupling could decrease the polarizabilities fluctuations of the buffer layer and therefore its Raman intensity. This phenomenon deserves more thorough theoretical investigations.

\bibliography{Biblio_Antoine.bib}

\begin{thebibliography}{37}%
\makeatletter
\providecommand \@ifxundefined [1]{%
 \@ifx{#1\undefined}
}%
\providecommand \@ifnum [1]{%
 \ifnum #1\expandafter \@firstoftwo
 \else \expandafter \@secondoftwo
 \fi
}%
\providecommand \@ifx [1]{%
 \ifx #1\expandafter \@firstoftwo
 \else \expandafter \@secondoftwo
 \fi
}%
\providecommand \natexlab [1]{#1}%
\providecommand \enquote  [1]{``#1''}%
\providecommand \bibnamefont  [1]{#1}%
\providecommand \bibfnamefont [1]{#1}%
\providecommand \citenamefont [1]{#1}%
\providecommand \href@noop [0]{\@secondoftwo}%
\providecommand \href [0]{\begingroup \@sanitize@url \@href}%
\providecommand \@href[1]{\@@startlink{#1}\@@href}%
\providecommand \@@href[1]{\endgroup#1\@@endlink}%
\providecommand \@sanitize@url [0]{\catcode `\\12\catcode `\$12\catcode
  `\&12\catcode `\#12\catcode `\^12\catcode `\_12\catcode `\%12\relax}%
\providecommand \@@startlink[1]{}%
\providecommand \@@endlink[0]{}%
\providecommand \url  [0]{\begingroup\@sanitize@url \@url }%
\providecommand \@url [1]{\endgroup\@href {#1}{\urlprefix }}%
\providecommand \urlprefix  [0]{URL }%
\providecommand \Eprint [0]{\href }%
\providecommand \doibase [0]{http://dx.doi.org/}%
\providecommand \selectlanguage [0]{\@gobble}%
\providecommand \bibinfo  [0]{\@secondoftwo}%
\providecommand \bibfield  [0]{\@secondoftwo}%
\providecommand \translation [1]{[#1]}%
\providecommand \BibitemOpen [0]{}%
\providecommand \bibitemStop [0]{}%
\providecommand \bibitemNoStop [0]{.\EOS\space}%
\providecommand \EOS [0]{\spacefactor3000\relax}%
\providecommand \BibitemShut  [1]{\csname bibitem#1\endcsname}%
\let\auto@bib@innerbib\@empty
\bibitem [{\citenamefont {Srivastava}\ \emph {et~al.}(2012)\citenamefont
  {Srivastava}, \citenamefont {He}, \citenamefont {Luxmi}, \citenamefont
  {Mende}, \citenamefont {Feenstra},\ and\ \citenamefont
  {Sun}}]{0022-3727-45-15-154001}%
  \BibitemOpen
  \bibfield  {author} {\bibinfo {author} {\bibfnamefont {N.}~\bibnamefont
  {Srivastava}}, \bibinfo {author} {\bibfnamefont {G.}~\bibnamefont {He}},
  \bibinfo {author} {\bibnamefont {Luxmi}}, \bibinfo {author} {\bibfnamefont
  {P.~C.}\ \bibnamefont {Mende}}, \bibinfo {author} {\bibfnamefont {R.~M.}\
  \bibnamefont {Feenstra}}, \ and\ \bibinfo {author} {\bibfnamefont
  {Y.}~\bibnamefont {Sun}},\ }\href
  {http://stacks.iop.org/0022-3727/45/i=15/a=154001} {\bibfield  {journal}
  {\bibinfo  {journal} {Journal of Physics D: Applied Physics}\ }\textbf
  {\bibinfo {volume} {45}},\ \bibinfo {pages} {154001} (\bibinfo {year}
  {2012})}\BibitemShut {NoStop}%
\bibitem [{\citenamefont {Berger}\ \emph {et~al.}(2004)\citenamefont {Berger},
  \citenamefont {Song}, \citenamefont {Li}, \citenamefont {Li}, \citenamefont
  {Ogbazghi}, \citenamefont {Feng}, \citenamefont {Dai}, \citenamefont
  {Marchenkov}, \citenamefont {Conrad}, \citenamefont {First},\ and\
  \citenamefont {de~Heer}}]{Berger:2004bh}%
  \BibitemOpen
  \bibfield  {author} {\bibinfo {author} {\bibfnamefont {C.}~\bibnamefont
  {Berger}}, \bibinfo {author} {\bibfnamefont {Z.}~\bibnamefont {Song}},
  \bibinfo {author} {\bibfnamefont {T.}~\bibnamefont {Li}}, \bibinfo {author}
  {\bibfnamefont {X.}~\bibnamefont {Li}}, \bibinfo {author} {\bibfnamefont
  {A.}~\bibnamefont {Ogbazghi}}, \bibinfo {author} {\bibfnamefont
  {R.}~\bibnamefont {Feng}}, \bibinfo {author} {\bibfnamefont {Z.}~\bibnamefont
  {Dai}}, \bibinfo {author} {\bibfnamefont {A.}~\bibnamefont {Marchenkov}},
  \bibinfo {author} {\bibfnamefont {E.}~\bibnamefont {Conrad}}, \bibinfo
  {author} {\bibfnamefont {P.}~\bibnamefont {First}}, \ and\ \bibinfo {author}
  {\bibfnamefont {W.}~\bibnamefont {de~Heer}},\ }\href {\doibase DOI
  10.1021/jp040650f} {\bibfield  {journal} {\bibinfo  {journal} {Journal of
  Physical Chemistry B}\ }\textbf {\bibinfo {volume} {108}},\ \bibinfo {pages}
  {19912} (\bibinfo {year} {2004})}\BibitemShut {NoStop}%
\bibitem [{\citenamefont {Sutter}(2009)}]{Sutter:2009dq}%
  \BibitemOpen
  \bibfield  {author} {\bibinfo {author} {\bibfnamefont {P.}~\bibnamefont
  {Sutter}},\ }\href {\doibase 10.1038/nmat2392} {\bibfield  {journal}
  {\bibinfo  {journal} {Nature Materials}\ }\textbf {\bibinfo {volume} {8}},\
  \bibinfo {pages} {171} (\bibinfo {year} {2009})}\BibitemShut {NoStop}%
\bibitem [{\citenamefont {Virojanadara}\ \emph {et~al.}(2008)\citenamefont
  {Virojanadara}, \citenamefont {Syv\"ajarvi}, \citenamefont {Yakimova},
  \citenamefont {Johansson}, \citenamefont {Zakharov},\ and\ \citenamefont
  {Balasubramanian}}]{PhysRevB.78.245403}%
  \BibitemOpen
  \bibfield  {author} {\bibinfo {author} {\bibfnamefont {C.}~\bibnamefont
  {Virojanadara}}, \bibinfo {author} {\bibfnamefont {M.}~\bibnamefont
  {Syv\"ajarvi}}, \bibinfo {author} {\bibfnamefont {R.}~\bibnamefont
  {Yakimova}}, \bibinfo {author} {\bibfnamefont {L.~I.}\ \bibnamefont
  {Johansson}}, \bibinfo {author} {\bibfnamefont {A.~A.}\ \bibnamefont
  {Zakharov}}, \ and\ \bibinfo {author} {\bibfnamefont {T.}~\bibnamefont
  {Balasubramanian}},\ }\href {\doibase 10.1103/PhysRevB.78.245403} {\bibfield
  {journal} {\bibinfo  {journal} {Physical Review B}\ }\textbf {\bibinfo
  {volume} {78}},\ \bibinfo {pages} {245403} (\bibinfo {year}
  {2008})}\BibitemShut {NoStop}%
\bibitem [{\citenamefont {Emtsev}\ \emph {et~al.}(2009)\citenamefont {Emtsev},
  \citenamefont {Bostwick}, \citenamefont {Horn}, \citenamefont {Jobst},
  \citenamefont {Kellogg}, \citenamefont {Ley}, \citenamefont {McChesney},
  \citenamefont {Ohta}, \citenamefont {Reshanov}, \citenamefont {Roehrl},
  \citenamefont {Rotenberg}, \citenamefont {Schmid}, \citenamefont {Waldmann},
  \citenamefont {Weber},\ and\ \citenamefont {Seyller}}]{Emtsev:2009nx}%
  \BibitemOpen
  \bibfield  {author} {\bibinfo {author} {\bibfnamefont {K.~V.}\ \bibnamefont
  {Emtsev}}, \bibinfo {author} {\bibfnamefont {A.}~\bibnamefont {Bostwick}},
  \bibinfo {author} {\bibfnamefont {K.}~\bibnamefont {Horn}}, \bibinfo {author}
  {\bibfnamefont {J.}~\bibnamefont {Jobst}}, \bibinfo {author} {\bibfnamefont
  {G.~L.}\ \bibnamefont {Kellogg}}, \bibinfo {author} {\bibfnamefont
  {L.}~\bibnamefont {Ley}}, \bibinfo {author} {\bibfnamefont {J.~L.}\
  \bibnamefont {McChesney}}, \bibinfo {author} {\bibfnamefont {T.}~\bibnamefont
  {Ohta}}, \bibinfo {author} {\bibfnamefont {S.~A.}\ \bibnamefont {Reshanov}},
  \bibinfo {author} {\bibfnamefont {J.}~\bibnamefont {Roehrl}}, \bibinfo
  {author} {\bibfnamefont {E.}~\bibnamefont {Rotenberg}}, \bibinfo {author}
  {\bibfnamefont {A.~K.}\ \bibnamefont {Schmid}}, \bibinfo {author}
  {\bibfnamefont {D.}~\bibnamefont {Waldmann}}, \bibinfo {author}
  {\bibfnamefont {H.~B.}\ \bibnamefont {Weber}}, \ and\ \bibinfo {author}
  {\bibfnamefont {T.}~\bibnamefont {Seyller}},\ }\href {\doibase DOI
  10.1038/NMAT2382} {\bibfield  {journal} {\bibinfo  {journal} {Nature
  Materials}\ }\textbf {\bibinfo {volume} {8}},\ \bibinfo {pages} {203}
  (\bibinfo {year} {2009})}\BibitemShut {NoStop}%
\bibitem [{\citenamefont {Tromp}\ and\ \citenamefont
  {Hannon}(2009)}]{PhysRevLett.102.106104}%
  \BibitemOpen
  \bibfield  {author} {\bibinfo {author} {\bibfnamefont {R.~M.}\ \bibnamefont
  {Tromp}}\ and\ \bibinfo {author} {\bibfnamefont {J.~B.}\ \bibnamefont
  {Hannon}},\ }\href {\doibase 10.1103/PhysRevLett.102.106104} {\bibfield
  {journal} {\bibinfo  {journal} {Phys. Rev. Lett.}\ }\textbf {\bibinfo
  {volume} {102}},\ \bibinfo {pages} {106104} (\bibinfo {year}
  {2009})}\BibitemShut {NoStop}%
\bibitem [{\citenamefont {Kim}\ \emph {et~al.}(2008)\citenamefont {Kim},
  \citenamefont {Ihm}, \citenamefont {Choi},\ and\ \citenamefont
  {Son}}]{PhysRevLett.100.176802}%
  \BibitemOpen
  \bibfield  {author} {\bibinfo {author} {\bibfnamefont {S.}~\bibnamefont
  {Kim}}, \bibinfo {author} {\bibfnamefont {J.}~\bibnamefont {Ihm}}, \bibinfo
  {author} {\bibfnamefont {H.~J.}\ \bibnamefont {Choi}}, \ and\ \bibinfo
  {author} {\bibfnamefont {Y.-W.}\ \bibnamefont {Son}},\ }\href {\doibase
  10.1103/PhysRevLett.100.176802} {\bibfield  {journal} {\bibinfo  {journal}
  {Physical Review Letters}\ }\textbf {\bibinfo {volume} {100}},\ \bibinfo
  {pages} {176802} (\bibinfo {year} {2008})}\BibitemShut {NoStop}%
\bibitem [{\citenamefont {Qi}\ \emph {et~al.}(2010)\citenamefont {Qi},
  \citenamefont {Rhim}, \citenamefont {Sun}, \citenamefont {Weinert},\ and\
  \citenamefont {Li}}]{PhysRevLett.105.085502}%
  \BibitemOpen
  \bibfield  {author} {\bibinfo {author} {\bibfnamefont {Y.}~\bibnamefont
  {Qi}}, \bibinfo {author} {\bibfnamefont {S.~H.}\ \bibnamefont {Rhim}},
  \bibinfo {author} {\bibfnamefont {G.~F.}\ \bibnamefont {Sun}}, \bibinfo
  {author} {\bibfnamefont {M.}~\bibnamefont {Weinert}}, \ and\ \bibinfo
  {author} {\bibfnamefont {L.}~\bibnamefont {Li}},\ }\href {\doibase
  10.1103/PhysRevLett.105.085502} {\bibfield  {journal} {\bibinfo  {journal}
  {Physical Review Letters}\ }\textbf {\bibinfo {volume} {105}},\ \bibinfo
  {pages} {085502} (\bibinfo {year} {2010})}\BibitemShut {NoStop}%
\bibitem [{\citenamefont {Varchon}\ \emph {et~al.}(2008)\citenamefont
  {Varchon}, \citenamefont {Mallet}, \citenamefont {Veuillen},\ and\
  \citenamefont {Magaud}}]{PhysRevB.77.235412}%
  \BibitemOpen
  \bibfield  {author} {\bibinfo {author} {\bibfnamefont {F.}~\bibnamefont
  {Varchon}}, \bibinfo {author} {\bibfnamefont {P.}~\bibnamefont {Mallet}},
  \bibinfo {author} {\bibfnamefont {J.-Y.}\ \bibnamefont {Veuillen}}, \ and\
  \bibinfo {author} {\bibfnamefont {L.}~\bibnamefont {Magaud}},\ }\href
  {\doibase 10.1103/PhysRevB.77.235412} {\bibfield  {journal} {\bibinfo
  {journal} {Physical Review B}\ }\textbf {\bibinfo {volume} {77}},\ \bibinfo
  {pages} {235412} (\bibinfo {year} {2008})}\BibitemShut {NoStop}%
\bibitem [{\citenamefont {Hoster}\ \emph {et~al.}(1997)\citenamefont {Hoster},
  \citenamefont {Kulakov},\ and\ \citenamefont {Bullemer}}]{Hoster:1997tg}%
  \BibitemOpen
  \bibfield  {author} {\bibinfo {author} {\bibfnamefont {H.}~\bibnamefont
  {Hoster}}, \bibinfo {author} {\bibfnamefont {M.}~\bibnamefont {Kulakov}}, \
  and\ \bibinfo {author} {\bibfnamefont {B.}~\bibnamefont {Bullemer}},\
  }\href@noop {} {\bibfield  {journal} {\bibinfo  {journal} {Surface Science}\
  }\textbf {\bibinfo {volume} {382}},\ \bibinfo {pages} {L658} (\bibinfo {year}
  {1997})}\BibitemShut {NoStop}%
\bibitem [{\citenamefont {Emtsev}\ \emph {et~al.}(2008)\citenamefont {Emtsev},
  \citenamefont {Speck}, \citenamefont {Seyller}, \citenamefont {Ley},\ and\
  \citenamefont {Riley}}]{PhysRevB.77.155303}%
  \BibitemOpen
  \bibfield  {author} {\bibinfo {author} {\bibfnamefont {K.~V.}\ \bibnamefont
  {Emtsev}}, \bibinfo {author} {\bibfnamefont {F.}~\bibnamefont {Speck}},
  \bibinfo {author} {\bibfnamefont {T.}~\bibnamefont {Seyller}}, \bibinfo
  {author} {\bibfnamefont {L.}~\bibnamefont {Ley}}, \ and\ \bibinfo {author}
  {\bibfnamefont {J.~D.}\ \bibnamefont {Riley}},\ }\href {\doibase
  10.1103/PhysRevB.77.155303} {\bibfield  {journal} {\bibinfo  {journal} {Phys.
  Rev. B}\ }\textbf {\bibinfo {volume} {77}},\ \bibinfo {pages} {155303}
  (\bibinfo {year} {2008})}\BibitemShut {NoStop}%
\bibitem [{\citenamefont {Hiebel}\ \emph {et~al.}(2008)\citenamefont {Hiebel},
  \citenamefont {Mallet}, \citenamefont {Varchon}, \citenamefont {Magaud},\
  and\ \citenamefont {Veuillen}}]{PhysRevB.78.153412}%
  \BibitemOpen
  \bibfield  {author} {\bibinfo {author} {\bibfnamefont {F.}~\bibnamefont
  {Hiebel}}, \bibinfo {author} {\bibfnamefont {P.}~\bibnamefont {Mallet}},
  \bibinfo {author} {\bibfnamefont {F.}~\bibnamefont {Varchon}}, \bibinfo
  {author} {\bibfnamefont {L.}~\bibnamefont {Magaud}}, \ and\ \bibinfo {author}
  {\bibfnamefont {J.-Y.}\ \bibnamefont {Veuillen}},\ }\href {\doibase
  10.1103/PhysRevB.78.153412} {\bibfield  {journal} {\bibinfo  {journal} {Phys.
  Rev. B}\ }\textbf {\bibinfo {volume} {78}},\ \bibinfo {pages} {153412}
  (\bibinfo {year} {2008})}\BibitemShut {NoStop}%
\bibitem [{\citenamefont {Hiebel}\ \emph {et~al.}(2009)\citenamefont {Hiebel},
  \citenamefont {Mallet}, \citenamefont {Magaud},\ and\ \citenamefont
  {Veuillen}}]{PhysRevB.80.235429}%
  \BibitemOpen
  \bibfield  {author} {\bibinfo {author} {\bibfnamefont {F.}~\bibnamefont
  {Hiebel}}, \bibinfo {author} {\bibfnamefont {P.}~\bibnamefont {Mallet}},
  \bibinfo {author} {\bibfnamefont {L.}~\bibnamefont {Magaud}}, \ and\ \bibinfo
  {author} {\bibfnamefont {J.-Y.}\ \bibnamefont {Veuillen}},\ }\href {\doibase
  10.1103/PhysRevB.80.235429} {\bibfield  {journal} {\bibinfo  {journal} {Phys.
  Rev. B}\ }\textbf {\bibinfo {volume} {80}},\ \bibinfo {pages} {235429}
  (\bibinfo {year} {2009})}\BibitemShut {NoStop}%
\bibitem [{\citenamefont {Berger}\ \emph {et~al.}(2006)\citenamefont {Berger},
  \citenamefont {Song}, \citenamefont {Li}, \citenamefont {Wu}, \citenamefont
  {Brown}, \citenamefont {Naud}, \citenamefont {Mayou}, \citenamefont {Li},
  \citenamefont {Hass}, \citenamefont {Marchenkov}, \citenamefont {Conrad},
  \citenamefont {First},\ and\ \citenamefont {de~Heer}}]{Berger:2006qf}%
  \BibitemOpen
  \bibfield  {author} {\bibinfo {author} {\bibfnamefont {C.}~\bibnamefont
  {Berger}}, \bibinfo {author} {\bibfnamefont {Z.}~\bibnamefont {Song}},
  \bibinfo {author} {\bibfnamefont {X.}~\bibnamefont {Li}}, \bibinfo {author}
  {\bibfnamefont {X.}~\bibnamefont {Wu}}, \bibinfo {author} {\bibfnamefont
  {N.}~\bibnamefont {Brown}}, \bibinfo {author} {\bibfnamefont
  {C.}~\bibnamefont {Naud}}, \bibinfo {author} {\bibfnamefont {D.}~\bibnamefont
  {Mayou}}, \bibinfo {author} {\bibfnamefont {T.}~\bibnamefont {Li}}, \bibinfo
  {author} {\bibfnamefont {J.}~\bibnamefont {Hass}}, \bibinfo {author}
  {\bibfnamefont {A.~N.}\ \bibnamefont {Marchenkov}}, \bibinfo {author}
  {\bibfnamefont {E.~H.}\ \bibnamefont {Conrad}}, \bibinfo {author}
  {\bibfnamefont {P.~N.}\ \bibnamefont {First}}, \ and\ \bibinfo {author}
  {\bibfnamefont {W.~A.}\ \bibnamefont {de~Heer}},\ }\href {\doibase
  10.1126/science.1125925} {\bibfield  {journal} {\bibinfo  {journal}
  {Science}\ }\textbf {\bibinfo {volume} {312}},\ \bibinfo {pages} {1191}
  (\bibinfo {year} {2006})}\BibitemShut {NoStop}%
\bibitem [{\citenamefont {Wu}\ \emph {et~al.}(2009)\citenamefont {Wu},
  \citenamefont {Hu}, \citenamefont {Ruan}, \citenamefont {Madiomanana},
  \citenamefont {Hankinson}, \citenamefont {Sprinkle}, \citenamefont {Berger},\
  and\ \citenamefont {de~Heer}}]{wu:223108}%
  \BibitemOpen
  \bibfield  {author} {\bibinfo {author} {\bibfnamefont {X.}~\bibnamefont
  {Wu}}, \bibinfo {author} {\bibfnamefont {Y.}~\bibnamefont {Hu}}, \bibinfo
  {author} {\bibfnamefont {M.}~\bibnamefont {Ruan}}, \bibinfo {author}
  {\bibfnamefont {N.~K.}\ \bibnamefont {Madiomanana}}, \bibinfo {author}
  {\bibfnamefont {J.}~\bibnamefont {Hankinson}}, \bibinfo {author}
  {\bibfnamefont {M.}~\bibnamefont {Sprinkle}}, \bibinfo {author}
  {\bibfnamefont {C.}~\bibnamefont {Berger}}, \ and\ \bibinfo {author}
  {\bibfnamefont {W.~A.}\ \bibnamefont {de~Heer}},\ }\href {\doibase
  10.1063/1.3266524} {\bibfield  {journal} {\bibinfo  {journal} {Applied
  Physics Letters}\ }\textbf {\bibinfo {volume} {95}},\ \bibinfo {eid} {223108}
  (\bibinfo {year} {2009})}\BibitemShut {NoStop}%
\bibitem [{\citenamefont {Varchon}\ \emph {et~al.}(2007)\citenamefont
  {Varchon}, \citenamefont {Feng}, \citenamefont {Hass}, \citenamefont {Li},
  \citenamefont {Nguyen}, \citenamefont {Naud}, \citenamefont {Mallet},
  \citenamefont {Veuillen}, \citenamefont {Berger}, \citenamefont {Conrad},\
  and\ \citenamefont {Magaud}}]{PhysRevLett.99.126805}%
  \BibitemOpen
  \bibfield  {author} {\bibinfo {author} {\bibfnamefont {F.}~\bibnamefont
  {Varchon}}, \bibinfo {author} {\bibfnamefont {R.}~\bibnamefont {Feng}},
  \bibinfo {author} {\bibfnamefont {J.}~\bibnamefont {Hass}}, \bibinfo {author}
  {\bibfnamefont {X.}~\bibnamefont {Li}}, \bibinfo {author} {\bibfnamefont
  {B.~N.}\ \bibnamefont {Nguyen}}, \bibinfo {author} {\bibfnamefont
  {C.}~\bibnamefont {Naud}}, \bibinfo {author} {\bibfnamefont {P.}~\bibnamefont
  {Mallet}}, \bibinfo {author} {\bibfnamefont {J.-Y.}\ \bibnamefont
  {Veuillen}}, \bibinfo {author} {\bibfnamefont {C.}~\bibnamefont {Berger}},
  \bibinfo {author} {\bibfnamefont {E.~H.}\ \bibnamefont {Conrad}}, \ and\
  \bibinfo {author} {\bibfnamefont {L.}~\bibnamefont {Magaud}},\ }\href
  {\doibase 10.1103/PhysRevLett.99.126805} {\bibfield  {journal} {\bibinfo
  {journal} {Phys. Rev. Lett.}\ }\textbf {\bibinfo {volume} {99}},\ \bibinfo
  {pages} {126805} (\bibinfo {year} {2007})}\BibitemShut {NoStop}%
\bibitem [{\citenamefont {Riedl}\ \emph {et~al.}(2010)\citenamefont {Riedl},
  \citenamefont {Coletti},\ and\ \citenamefont
  {Starke}}]{0022-3727-43-37-374009}%
  \BibitemOpen
  \bibfield  {author} {\bibinfo {author} {\bibfnamefont {C.}~\bibnamefont
  {Riedl}}, \bibinfo {author} {\bibfnamefont {C.}~\bibnamefont {Coletti}}, \
  and\ \bibinfo {author} {\bibfnamefont {U.}~\bibnamefont {Starke}},\ }\href
  {http://stacks.iop.org/0022-3727/43/i=37/a=374009} {\bibfield  {journal}
  {\bibinfo  {journal} {Journal of Physics D: Applied Physics}\ }\textbf
  {\bibinfo {volume} {43}},\ \bibinfo {pages} {374009} (\bibinfo {year}
  {2010})}\BibitemShut {NoStop}%
\bibitem [{\citenamefont {Virojanadara}\ \emph
  {et~al.}(2010{\natexlab{a}})\citenamefont {Virojanadara}, \citenamefont
  {Zakharov}, \citenamefont {Yakimova},\ and\ \citenamefont
  {Johansson}}]{Virojanadara2010L4}%
  \BibitemOpen
  \bibfield  {author} {\bibinfo {author} {\bibfnamefont {C.}~\bibnamefont
  {Virojanadara}}, \bibinfo {author} {\bibfnamefont {A.}~\bibnamefont
  {Zakharov}}, \bibinfo {author} {\bibfnamefont {R.}~\bibnamefont {Yakimova}},
  \ and\ \bibinfo {author} {\bibfnamefont {L.}~\bibnamefont {Johansson}},\
  }\href {\doibase 10.1016/j.susc.2009.11.011} {\bibfield  {journal} {\bibinfo
  {journal} {Surface Science}\ }\textbf {\bibinfo {volume} {604}},\ \bibinfo
  {pages} {L4 } (\bibinfo {year} {2010}{\natexlab{a}})}\BibitemShut {NoStop}%
\bibitem [{\citenamefont {Virojanadara}\ \emph
  {et~al.}(2010{\natexlab{b}})\citenamefont {Virojanadara}, \citenamefont
  {Yakimova}, \citenamefont {Zakharov},\ and\ \citenamefont
  {Johansson}}]{0022-3727-43-37-374010}%
  \BibitemOpen
  \bibfield  {author} {\bibinfo {author} {\bibfnamefont {C.}~\bibnamefont
  {Virojanadara}}, \bibinfo {author} {\bibfnamefont {R.}~\bibnamefont
  {Yakimova}}, \bibinfo {author} {\bibfnamefont {A.~A.}\ \bibnamefont
  {Zakharov}}, \ and\ \bibinfo {author} {\bibfnamefont {L.~I.}\ \bibnamefont
  {Johansson}},\ }\href {http://stacks.iop.org/0022-3727/43/i=37/a=374010}
  {\bibfield  {journal} {\bibinfo  {journal} {Journal of Physics D: Applied
  Physics}\ }\textbf {\bibinfo {volume} {43}},\ \bibinfo {pages} {374010}
  (\bibinfo {year} {2010}{\natexlab{b}})}\BibitemShut {NoStop}%
\bibitem [{\citenamefont {Riedl}\ \emph {et~al.}(2009)\citenamefont {Riedl},
  \citenamefont {Coletti}, \citenamefont {Iwasaki}, \citenamefont {Zakharov},\
  and\ \citenamefont {Starke}}]{PhysRevLett.103.246804}%
  \BibitemOpen
  \bibfield  {author} {\bibinfo {author} {\bibfnamefont {C.}~\bibnamefont
  {Riedl}}, \bibinfo {author} {\bibfnamefont {C.}~\bibnamefont {Coletti}},
  \bibinfo {author} {\bibfnamefont {T.}~\bibnamefont {Iwasaki}}, \bibinfo
  {author} {\bibfnamefont {A.~A.}\ \bibnamefont {Zakharov}}, \ and\ \bibinfo
  {author} {\bibfnamefont {U.}~\bibnamefont {Starke}},\ }\href {\doibase
  10.1103/PhysRevLett.103.246804} {\bibfield  {journal} {\bibinfo  {journal}
  {Phys. Rev. Lett.}\ }\textbf {\bibinfo {volume} {103}},\ \bibinfo {pages}
  {246804} (\bibinfo {year} {2009})}\BibitemShut {NoStop}%
\bibitem [{\citenamefont {Forti}\ \emph {et~al.}(2011)\citenamefont {Forti},
  \citenamefont {Emtsev}, \citenamefont {Coletti}, \citenamefont {Zakharov},
  \citenamefont {Riedl},\ and\ \citenamefont {Starke}}]{PhysRevB.84.125449}%
  \BibitemOpen
  \bibfield  {author} {\bibinfo {author} {\bibfnamefont {S.}~\bibnamefont
  {Forti}}, \bibinfo {author} {\bibfnamefont {K.~V.}\ \bibnamefont {Emtsev}},
  \bibinfo {author} {\bibfnamefont {C.}~\bibnamefont {Coletti}}, \bibinfo
  {author} {\bibfnamefont {A.~A.}\ \bibnamefont {Zakharov}}, \bibinfo {author}
  {\bibfnamefont {C.}~\bibnamefont {Riedl}}, \ and\ \bibinfo {author}
  {\bibfnamefont {U.}~\bibnamefont {Starke}},\ }\href {\doibase
  10.1103/PhysRevB.84.125449} {\bibfield  {journal} {\bibinfo  {journal} {Phys.
  Rev. B}\ }\textbf {\bibinfo {volume} {84}},\ \bibinfo {pages} {125449}
  (\bibinfo {year} {2011})}\BibitemShut {NoStop}%
\bibitem [{\citenamefont {Starke}\ and\ \citenamefont
  {Riedl}(2009)}]{0953-8984-21-13-134016}%
  \BibitemOpen
  \bibfield  {author} {\bibinfo {author} {\bibfnamefont {U.}~\bibnamefont
  {Starke}}\ and\ \bibinfo {author} {\bibfnamefont {C.}~\bibnamefont {Riedl}},\
  }\href {http://stacks.iop.org/0953-8984/21/i=13/a=134016} {\bibfield
  {journal} {\bibinfo  {journal} {Journal of Physics: Condensed Matter}\
  }\textbf {\bibinfo {volume} {21}},\ \bibinfo {pages} {134016} (\bibinfo
  {year} {2009})}\BibitemShut {NoStop}%
\bibitem [{\citenamefont {Michon}\ \emph {et~al.}(2010)\citenamefont {Michon},
  \citenamefont {V{\'e}zian}, \citenamefont {Ouerghi}, \citenamefont
  {Zielinski}, \citenamefont {Chassagne},\ and\ \citenamefont
  {Portail}}]{10.1063/1.3503972}%
  \BibitemOpen
  \bibfield  {author} {\bibinfo {author} {\bibfnamefont {A.}~\bibnamefont
  {Michon}}, \bibinfo {author} {\bibfnamefont {S.}~\bibnamefont {V{\'e}zian}},
  \bibinfo {author} {\bibfnamefont {A.}~\bibnamefont {Ouerghi}}, \bibinfo
  {author} {\bibfnamefont {M.}~\bibnamefont {Zielinski}}, \bibinfo {author}
  {\bibfnamefont {T.}~\bibnamefont {Chassagne}}, \ and\ \bibinfo {author}
  {\bibfnamefont {M.}~\bibnamefont {Portail}},\ }\href {\doibase
  DOI:10.1063/1.3503972} {\bibfield  {journal} {\bibinfo  {journal} {Appl.
  Phys. Lett.}\ }\textbf {\bibinfo {volume} {97}},\ \bibinfo {pages} {171909}
  (\bibinfo {year} {2010})}\BibitemShut {NoStop}%
\bibitem [{\citenamefont {Ferralis}\ \emph {et~al.}(2008)\citenamefont
  {Ferralis}, \citenamefont {Maboudian},\ and\ \citenamefont
  {Carraro}}]{PhysRevLett.101.156801}%
  \BibitemOpen
  \bibfield  {author} {\bibinfo {author} {\bibfnamefont {N.}~\bibnamefont
  {Ferralis}}, \bibinfo {author} {\bibfnamefont {R.}~\bibnamefont {Maboudian}},
  \ and\ \bibinfo {author} {\bibfnamefont {C.}~\bibnamefont {Carraro}},\ }\href
  {\doibase 10.1103/PhysRevLett.101.156801} {\bibfield  {journal} {\bibinfo
  {journal} {Phys. Rev. Lett.}\ }\textbf {\bibinfo {volume} {101}},\ \bibinfo
  {pages} {156801} (\bibinfo {year} {2008})}\BibitemShut {NoStop}%
\bibitem [{\citenamefont {Camara}\ \emph {et~al.}(2009)\citenamefont {Camara},
  \citenamefont {Huntzinger}, \citenamefont {Rius}, \citenamefont {Tiberj},
  \citenamefont {Mestres}, \citenamefont {P\'erez-Murano}, \citenamefont
  {Godignon},\ and\ \citenamefont {Camassel}}]{PhysRevB.80.125410}%
  \BibitemOpen
  \bibfield  {author} {\bibinfo {author} {\bibfnamefont {N.}~\bibnamefont
  {Camara}}, \bibinfo {author} {\bibfnamefont {J.-R.}\ \bibnamefont
  {Huntzinger}}, \bibinfo {author} {\bibfnamefont {G.}~\bibnamefont {Rius}},
  \bibinfo {author} {\bibfnamefont {A.}~\bibnamefont {Tiberj}}, \bibinfo
  {author} {\bibfnamefont {N.}~\bibnamefont {Mestres}}, \bibinfo {author}
  {\bibfnamefont {F.}~\bibnamefont {P\'erez-Murano}}, \bibinfo {author}
  {\bibfnamefont {P.}~\bibnamefont {Godignon}}, \ and\ \bibinfo {author}
  {\bibfnamefont {J.}~\bibnamefont {Camassel}},\ }\href {\doibase
  10.1103/PhysRevB.80.125410} {\bibfield  {journal} {\bibinfo  {journal} {Phys.
  Rev. B}\ }\textbf {\bibinfo {volume} {80}},\ \bibinfo {pages} {125410}
  (\bibinfo {year} {2009})}\BibitemShut {NoStop}%
\bibitem [{\citenamefont {Camara}\ \emph {et~al.}(2010)\citenamefont {Camara},
  \citenamefont {Tiberj}, \citenamefont {Jouault}, \citenamefont {Caboni},
  \citenamefont {Jabakhanji}, \citenamefont {Mestres}, \citenamefont
  {Godignon},\ and\ \citenamefont {Camassel}}]{Camara_2010}%
  \BibitemOpen
  \bibfield  {author} {\bibinfo {author} {\bibfnamefont {N.}~\bibnamefont
  {Camara}}, \bibinfo {author} {\bibfnamefont {A.}~\bibnamefont {Tiberj}},
  \bibinfo {author} {\bibfnamefont {B.}~\bibnamefont {Jouault}}, \bibinfo
  {author} {\bibfnamefont {A.}~\bibnamefont {Caboni}}, \bibinfo {author}
  {\bibfnamefont {B.}~\bibnamefont {Jabakhanji}}, \bibinfo {author}
  {\bibfnamefont {N.}~\bibnamefont {Mestres}}, \bibinfo {author} {\bibfnamefont
  {P.}~\bibnamefont {Godignon}}, \ and\ \bibinfo {author} {\bibfnamefont
  {J.}~\bibnamefont {Camassel}},\ }\href
  {http://stacks.iop.org/0022-3727/43/i=37/a=374011} {\bibfield  {journal}
  {\bibinfo  {journal} {J. Phys. D: Appl. Phys.}\ }\textbf {\bibinfo {volume}
  {43}},\ \bibinfo {pages} {374011} (\bibinfo {year} {2010})}\BibitemShut
  {NoStop}%
\bibitem [{\citenamefont {Burton}\ \emph {et~al.}(1999)\citenamefont {Burton},
  \citenamefont {Sun}, \citenamefont {Long}, \citenamefont {Feng},\ and\
  \citenamefont {Ferguson}}]{PhysRevB.59.7282}%
  \BibitemOpen
  \bibfield  {author} {\bibinfo {author} {\bibfnamefont {J.~C.}\ \bibnamefont
  {Burton}}, \bibinfo {author} {\bibfnamefont {L.}~\bibnamefont {Sun}},
  \bibinfo {author} {\bibfnamefont {F.~H.}\ \bibnamefont {Long}}, \bibinfo
  {author} {\bibfnamefont {Z.~C.}\ \bibnamefont {Feng}}, \ and\ \bibinfo
  {author} {\bibfnamefont {I.~T.}\ \bibnamefont {Ferguson}},\ }\href {\doibase
  10.1103/PhysRevB.59.7282} {\bibfield  {journal} {\bibinfo  {journal} {Phys.
  Rev. B}\ }\textbf {\bibinfo {volume} {59}},\ \bibinfo {pages} {7282}
  (\bibinfo {year} {1999})}\BibitemShut {NoStop}%
\bibitem [{\citenamefont {Englert}\ \emph {et~al.}(2011)\citenamefont
  {Englert}, \citenamefont {Dotzer}, \citenamefont {Yang}, \citenamefont
  {Schmid}, \citenamefont {Papp}, \citenamefont {Gottfried}, \citenamefont
  {Steinr{\~A}¼ck}, \citenamefont {Spiecker}, \citenamefont {Hauke},\ and\
  \citenamefont {Hirsch}}]{Englert:2011fk}%
  \BibitemOpen
  \bibfield  {author} {\bibinfo {author} {\bibfnamefont {J.~M.}\ \bibnamefont
  {Englert}}, \bibinfo {author} {\bibfnamefont {C.}~\bibnamefont {Dotzer}},
  \bibinfo {author} {\bibfnamefont {G.}~\bibnamefont {Yang}}, \bibinfo {author}
  {\bibfnamefont {M.}~\bibnamefont {Schmid}}, \bibinfo {author} {\bibfnamefont
  {C.}~\bibnamefont {Papp}}, \bibinfo {author} {\bibfnamefont {J.~M.}\
  \bibnamefont {Gottfried}}, \bibinfo {author} {\bibfnamefont {H.-P.}\
  \bibnamefont {Steinr{\~A}¼ck}}, \bibinfo {author} {\bibfnamefont
  {E.}~\bibnamefont {Spiecker}}, \bibinfo {author} {\bibfnamefont
  {F.}~\bibnamefont {Hauke}}, \ and\ \bibinfo {author} {\bibfnamefont
  {A.}~\bibnamefont {Hirsch}},\ }\href {http://dx.doi.org/10.1038/nchem.1010}
  {\bibfield  {journal} {\bibinfo  {journal} {Nat Chem}\ }\textbf {\bibinfo
  {volume} {3}},\ \bibinfo {pages} {279} (\bibinfo {year} {2011})}\BibitemShut
  {NoStop}%
\bibitem [{\citenamefont {Ferrari}\ and\ \citenamefont
  {Robertson}(2001)}]{PhysRevB.63.121405}%
  \BibitemOpen
  \bibfield  {author} {\bibinfo {author} {\bibfnamefont {A.~C.}\ \bibnamefont
  {Ferrari}}\ and\ \bibinfo {author} {\bibfnamefont {J.}~\bibnamefont
  {Robertson}},\ }\href {\doibase 10.1103/PhysRevB.63.121405} {\bibfield
  {journal} {\bibinfo  {journal} {Phys. Rev. B}\ }\textbf {\bibinfo {volume}
  {63}},\ \bibinfo {pages} {121405} (\bibinfo {year} {2001})}\BibitemShut
  {NoStop}%
\bibitem [{\citenamefont {Martins~Ferreira}\ \emph {et~al.}(2010)\citenamefont
  {Martins~Ferreira}, \citenamefont {Moutinho}, \citenamefont {Stavale},
  \citenamefont {Lucchese}, \citenamefont {Capaz}, \citenamefont {Achete},\
  and\ \citenamefont {Jorio}}]{PhysRevB.82.125429}%
  \BibitemOpen
  \bibfield  {author} {\bibinfo {author} {\bibfnamefont {E.~H.}\ \bibnamefont
  {Martins~Ferreira}}, \bibinfo {author} {\bibfnamefont {M.~V.~O.}\
  \bibnamefont {Moutinho}}, \bibinfo {author} {\bibfnamefont {F.}~\bibnamefont
  {Stavale}}, \bibinfo {author} {\bibfnamefont {M.~M.}\ \bibnamefont
  {Lucchese}}, \bibinfo {author} {\bibfnamefont {R.~B.}\ \bibnamefont {Capaz}},
  \bibinfo {author} {\bibfnamefont {C.~A.}\ \bibnamefont {Achete}}, \ and\
  \bibinfo {author} {\bibfnamefont {A.}~\bibnamefont {Jorio}},\ }\href
  {\doibase 10.1103/PhysRevB.82.125429} {\bibfield  {journal} {\bibinfo
  {journal} {Phys. Rev. B}\ }\textbf {\bibinfo {volume} {82}},\ \bibinfo
  {pages} {125429} (\bibinfo {year} {2010})}\BibitemShut {NoStop}%
\bibitem [{\citenamefont {Poncharal}\ \emph {et~al.}(2009)\citenamefont
  {Poncharal}, \citenamefont {Ayari}, \citenamefont {Michel},\ and\
  \citenamefont {Sauvajol}}]{PhysRevB.79.195417}%
  \BibitemOpen
  \bibfield  {author} {\bibinfo {author} {\bibfnamefont {P.}~\bibnamefont
  {Poncharal}}, \bibinfo {author} {\bibfnamefont {A.}~\bibnamefont {Ayari}},
  \bibinfo {author} {\bibfnamefont {T.}~\bibnamefont {Michel}}, \ and\ \bibinfo
  {author} {\bibfnamefont {J.-L.}\ \bibnamefont {Sauvajol}},\ }\href {\doibase
  10.1103/PhysRevB.79.195417} {\bibfield  {journal} {\bibinfo  {journal} {Phys.
  Rev. B}\ }\textbf {\bibinfo {volume} {79}},\ \bibinfo {pages} {195417}
  (\bibinfo {year} {2009})}\BibitemShut {NoStop}%
\bibitem [{\citenamefont {Ni}\ \emph {et~al.}(2008)\citenamefont {Ni},
  \citenamefont {Wang}, \citenamefont {Yu}, \citenamefont {You},\ and\
  \citenamefont {Shen}}]{PhysRevB.77.235403}%
  \BibitemOpen
  \bibfield  {author} {\bibinfo {author} {\bibfnamefont {Z.}~\bibnamefont
  {Ni}}, \bibinfo {author} {\bibfnamefont {Y.}~\bibnamefont {Wang}}, \bibinfo
  {author} {\bibfnamefont {T.}~\bibnamefont {Yu}}, \bibinfo {author}
  {\bibfnamefont {Y.}~\bibnamefont {You}}, \ and\ \bibinfo {author}
  {\bibfnamefont {Z.}~\bibnamefont {Shen}},\ }\href {\doibase
  10.1103/PhysRevB.77.235403} {\bibfield  {journal} {\bibinfo  {journal} {Phys.
  Rev. B}\ }\textbf {\bibinfo {volume} {77}},\ \bibinfo {pages} {235403}
  (\bibinfo {year} {2008})}\BibitemShut {NoStop}%
\bibitem [{\citenamefont {Tiberj}\ \emph {et~al.}(2011)\citenamefont {Tiberj},
  \citenamefont {Camara}, \citenamefont {Godignon},\ and\ \citenamefont
  {Camassel}}]{21801347}%
  \BibitemOpen
  \bibfield  {author} {\bibinfo {author} {\bibfnamefont {A.}~\bibnamefont
  {Tiberj}}, \bibinfo {author} {\bibfnamefont {N.}~\bibnamefont {Camara}},
  \bibinfo {author} {\bibfnamefont {P.}~\bibnamefont {Godignon}}, \ and\
  \bibinfo {author} {\bibfnamefont {J.}~\bibnamefont {Camassel}},\ }\href
  {\doibase 10.1186/1556-276X-6-478} {\bibfield  {journal} {\bibinfo  {journal}
  {Nanoscale Research Letters}\ }\textbf {\bibinfo {volume} {6}},\ \bibinfo
  {pages} {478} (\bibinfo {year} {2011})}\BibitemShut {NoStop}%
\bibitem [{\citenamefont {Benassi}\ \emph {et~al.}(1995)\citenamefont
  {Benassi}, \citenamefont {Frizzera}, \citenamefont {Montagna}, \citenamefont
  {Viliani}, \citenamefont {Mazzacurati}, \citenamefont {Ruocco},\ and\
  \citenamefont {Signorelli}}]{benassi95}%
  \BibitemOpen
  \bibfield  {author} {\bibinfo {author} {\bibfnamefont {P.}~\bibnamefont
  {Benassi}}, \bibinfo {author} {\bibfnamefont {W.}~\bibnamefont {Frizzera}},
  \bibinfo {author} {\bibfnamefont {M.}~\bibnamefont {Montagna}}, \bibinfo
  {author} {\bibfnamefont {G.}~\bibnamefont {Viliani}}, \bibinfo {author}
  {\bibfnamefont {V.}~\bibnamefont {Mazzacurati}}, \bibinfo {author}
  {\bibfnamefont {G.}~\bibnamefont {Ruocco}}, \ and\ \bibinfo {author}
  {\bibfnamefont {G.}~\bibnamefont {Signorelli}},\ }\href {\doibase
  10.1016/0378-4371(95)00055-C} {\bibfield  {journal} {\bibinfo  {journal}
  {Physica A: Statistical Mechanics and its Applications}\ }\textbf {\bibinfo
  {volume} {216}},\ \bibinfo {pages} {32} (\bibinfo {year} {1995})}\BibitemShut
  {NoStop}%
\bibitem [{\citenamefont {Berciaud}\ \emph {et~al.}(2009)\citenamefont
  {Berciaud}, \citenamefont {Ryu}, \citenamefont {Brus},\ and\ \citenamefont
  {Heinz}}]{doi:10.1021/nl8031444}%
  \BibitemOpen
  \bibfield  {author} {\bibinfo {author} {\bibfnamefont {S.}~\bibnamefont
  {Berciaud}}, \bibinfo {author} {\bibfnamefont {S.}~\bibnamefont {Ryu}},
  \bibinfo {author} {\bibfnamefont {L.~E.}\ \bibnamefont {Brus}}, \ and\
  \bibinfo {author} {\bibfnamefont {T.~F.}\ \bibnamefont {Heinz}},\ }\href
  {\doibase 10.1021/nl8031444} {\bibfield  {journal} {\bibinfo  {journal} {Nano
  Letters}\ }\textbf {\bibinfo {volume} {9}},\ \bibinfo {pages} {346} (\bibinfo
  {year} {2009})},\ \Eprint
  {http://arxiv.org/abs/http://pubs.acs.org/doi/pdf/10.1021/nl8031444}
  {http://pubs.acs.org/doi/pdf/10.1021/nl8031444} \BibitemShut {NoStop}%
\bibitem [{\citenamefont {Zabel}\ \emph {et~al.}(2012)\citenamefont {Zabel},
  \citenamefont {Nair}, \citenamefont {Ott}, \citenamefont {Georgiou},
  \citenamefont {Geim}, \citenamefont {Novoselov},\ and\ \citenamefont
  {Casiraghi}}]{doi:10.1021/nl203359n}%
  \BibitemOpen
  \bibfield  {author} {\bibinfo {author} {\bibfnamefont {J.}~\bibnamefont
  {Zabel}}, \bibinfo {author} {\bibfnamefont {R.~R.}\ \bibnamefont {Nair}},
  \bibinfo {author} {\bibfnamefont {A.}~\bibnamefont {Ott}}, \bibinfo {author}
  {\bibfnamefont {T.}~\bibnamefont {Georgiou}}, \bibinfo {author}
  {\bibfnamefont {A.~K.}\ \bibnamefont {Geim}}, \bibinfo {author}
  {\bibfnamefont {K.~S.}\ \bibnamefont {Novoselov}}, \ and\ \bibinfo {author}
  {\bibfnamefont {C.}~\bibnamefont {Casiraghi}},\ }\href {\doibase
  10.1021/nl203359n} {\bibfield  {journal} {\bibinfo  {journal} {Nano Letters}\
  }\textbf {\bibinfo {volume} {12}},\ \bibinfo {pages} {617} (\bibinfo {year}
  {2012})},\ \Eprint
  {http://arxiv.org/abs/http://pubs.acs.org/doi/pdf/10.1021/nl203359n}
  {http://pubs.acs.org/doi/pdf/10.1021/nl203359n} \BibitemShut {NoStop}%
\bibitem [{\citenamefont {Li}\ \emph {et~al.}(2009)\citenamefont {Li},
  \citenamefont {Cai}, \citenamefont {An}, \citenamefont {Kim}, \citenamefont
  {Nah}, \citenamefont {Yang}, \citenamefont {Piner}, \citenamefont
  {Velamakanni}, \citenamefont {Jung}, \citenamefont {Tutuc}, \citenamefont
  {Banerjee}, \citenamefont {Colombo},\ and\ \citenamefont
  {Ruoff}}]{Li:2009ly}%
  \BibitemOpen
  \bibfield  {author} {\bibinfo {author} {\bibfnamefont {X.}~\bibnamefont
  {Li}}, \bibinfo {author} {\bibfnamefont {W.}~\bibnamefont {Cai}}, \bibinfo
  {author} {\bibfnamefont {J.}~\bibnamefont {An}}, \bibinfo {author}
  {\bibfnamefont {S.}~\bibnamefont {Kim}}, \bibinfo {author} {\bibfnamefont
  {J.}~\bibnamefont {Nah}}, \bibinfo {author} {\bibfnamefont {D.}~\bibnamefont
  {Yang}}, \bibinfo {author} {\bibfnamefont {R.}~\bibnamefont {Piner}},
  \bibinfo {author} {\bibfnamefont {A.}~\bibnamefont {Velamakanni}}, \bibinfo
  {author} {\bibfnamefont {I.}~\bibnamefont {Jung}}, \bibinfo {author}
  {\bibfnamefont {E.}~\bibnamefont {Tutuc}}, \bibinfo {author} {\bibfnamefont
  {S.~K.}\ \bibnamefont {Banerjee}}, \bibinfo {author} {\bibfnamefont
  {L.}~\bibnamefont {Colombo}}, \ and\ \bibinfo {author} {\bibfnamefont
  {R.~S.}\ \bibnamefont {Ruoff}},\ }\href {\doibase 10.1126/science.1171245}
  {\bibfield  {journal} {\bibinfo  {journal} {Science}\ }\textbf {\bibinfo
  {volume} {324}},\ \bibinfo {pages} {1312} (\bibinfo {year}
  {2009})}\BibitemShut {NoStop}%
\end{thebibliography}%

\end{document}